\documentclass[12pt,twoside]{article}
\usepackage{amssymb}
\def\1#1{{\bf #1}}
\def\2#1{{\cal #1}}\def\3#1{{\sl #1}}\def\4#1{{\tt #1}}\def\5#1{{\sf #1}}
\def\6#1{{\mathfrak #1}}\def\7#1{{\mathbb #1}}\def\8#1{{\rm #1}}

\def\aut{{\rm Aut}}

\def\ol{\overline}
\def\ska{\vskip 0.2cm}

\def\beq{\begin{equation}}
\def\eeq{\end{equation}}
\def\vs{\vspace{0.2cm} \\}

\newtheorem{Def}{Definiton}[section]
\newtheorem{Lem}{Lemma}[section]
\newtheorem{The}{Theorem}[section]
\newtheorem{Pro}{Proposition}[section]
\newtheorem{Cor}{Corollary}[section]
\def\bdef{\begin{Def}\1: \em}
\def\eef{\end{Def}\ska}
\def\blem{\begin{Lem}\1: }
\def\elem{\end{Lem}\ska}
\def\bthe{\begin{The}\1: }
\def\ethe{\end{The}\ska}
\def\bpro{\begin{Pro}\1: }
\def\epro{\end{Pro}\ska}
\def\bcor{\begin{Cor}\1: }
\def\ecor{\end{Cor}\ska}
\def\supp{{\rm supp}}

\def\id{{\sf id}}

\def\al{\alpha}
\def\be{\beta}
\def\gam{\gamma}\def\Gam{\Gamma}
\def\lam{\lambda}\def\Lam{\Lambda}
\def\eps{\epsilon} 
\def\te{\theta}

\def\sgm{\sigma}\def\Sgm{\Sigma}
\def\om{\omega}\def\Om{\Omega}

\def\bpr{\noindent{\em Proof. }}

\def\epr{ $\square$\ska}

\def\pa{\partial}
\def\<{\langle}
\def\>{\rangle}
\def\Ad{{\sf Ad}}

\def\dyn{{\rm dyn}}
\def\ps{{\rm ps}}
\def\sec{{\rm sec}}
\def\bl{\biggl}
\def\br{\biggr}
\def\bm{\biggm}

\def\bdes{\begin{description}}
\def\edes{\end{description}}
\newcommand\itno[1]{\item[{\it (#1)}]}

\def\beqa{\begin{eqnarray*}}
\def\eeqa{\end{eqnarray*}}
\def\bdia{\begin{diagram}}
\def\edia{\end{diagram}}
\def\olt{\ \overline{\otimes} \ }
\def\bsl{\backslash}
\title{\bf
Kink States in $P(\phi)_2$-Models \\ (An Algebraic Approach)}
\author{{\it Dirk Schlingemann} \\
  II. Institut f\"ur Theoretische Physik \\
  Universit\"at Hamburg  \\ \\
 DESY 96-051}
\begin{document}
%%%%%%%%%%%%%%%%%%%%%%%%%%%%%%%%%%%%%%%%%%%%%%%%%%%%%%%%%%%%%%%%%%%%%%%%%
\maketitle
\abstract{
Several two-dimensional quantum field theory models 
have more than one vacuum state. Familiar examples are
the Sine-Gordon and the $\phi^4_2$-model. It is known 
that in these models there are also states, 
called kink states, which interpolate different vacua. 
A general construction scheme for 
kink states in the framework of algebraic quantum field theory is 
developed in a previous paper.
However, for the application of this method, the crucial condition is the  
split property for wedge algebras 
in the vacuum representations of the considered models. 
It is believed that the vacuum representations of 
$P(\phi)_2$-models fulfill this condition, but a 
rigorous proof is only known for the massive free scalar field. 
Therefore, we investigate in a  
construction of kink states which can directly be 
applied to $P(\phi)_2$-model, by making use of 
the properties of the dynamic of a $P(\phi)_2$-model.}
\thispagestyle{empty}
\newpage
%%%%%%%%%%%%%%%%%%%%%%%%%%%%%%%%%%%%%%%%%%%%%%%%%%%%%%%%%%%%%%%%%%%%%%%%%%%
%\tableofcontents   
%%%%%%%%%%%%%%%%%%%%%%%%%%%%%%%%%%%%%%%%%%%%%%%%%%%%%%%%%%%%%%%%%%%%%%%%%%
\section{Introduction and Overview}
%%%%%%%%%%%%%%%%%%%%%%%%%%%%%%%%%%%%%%%%%%%%%%%%%%%%%%%%%%%%%%%%%%%%%%%%%%%
There are familiar examples of $1+1$ dimensional quantum field theory models 
which possess more than one vacuum state. Let us mention 
the Sine-Gordon model, the $\phi^4_2$-theory and the Skyrme model. 
Further candidates are special types of $P(\phi)_2$-models.

It is known that the Sine-Gordon and the $\phi^4_2$-model possess 
states, called kink states, which interpolate different vacuum states. 
A construction of them was done 
by J. Fr\"ohlich in the 70s and 
can be obtained from \cite{Froh1}. 
In \cite[chapter 5]{Froh1}, J. Fr\"ohlich discusses the existence 
of kink states in general $P(\phi)_2$-models. 
However, this construction leads only to 
kink states which interpolate vacua which are connected 
by an (special) internal symmetry transformation, namely 
$\phi\mapsto -\phi$.
Moreover, a construction 
of the vacuum states of the $\phi^4_2$-model 
and their corresponding kink states is given in \cite{Froh3}
by using Euclidean methods.

We expect that there
are $P(\phi)_2$-models which have more than one vacuum state, but where these 
vacua are not related by an internal symmetry transformation.
For these purposes, we investigate the following question:

Let us consider a $1+1$-dimensional model of a quantum field theory 
which possesses more than one vacuum state, 
which conditions a pair of vacuum sates has to fulfill, such that an 
interpolating kink state can be constructed?

This question is already 
discussed in \cite{Schl4}, where a general construction
scheme for kink states is developed.
It is purely algebraic and independent of the specific properties of a model.
Another advantage is that 
the assumption that the vacua are related by an internal 
symmetry transformation, as is used in \cite{Froh1,Froh3}, is not needed.
On the other hand, to apply this construction scheme 
to a pair of vacuum states, the crucial condition is the {\em split property} 
for wedge algebras in the 
GNS-representations of the considered vacuum states. 
Hence we
have to prove this condition for pairs of vacuum states of the model 
under consideration if we want to apply these results to a concrete model. 
It is believed that the
vacuum states of $P(\phi)_2$-models fulfill this condition, but a 
rigorous proof is only known for the massive free scalar field 
\cite[Appendix of this paper]{AntFre,Bu1}. 

Therefore, we investigate 
a construction of kink states which can directly be 
applied to $P(\phi)_2$-models. 

We make use of the properties of the dynamic of a $P(\phi)_2$-model 
to show that  
the construction scheme, which is described in \cite{Schl4}, is also applicable
to $P(\phi)_2$-models.  
More precisely, it is sufficient to assume that the 
vacua under consideration have the local Fock property, which is 
automatically the case for each $P(\phi)_2$ vacuum \cite{GlJa1},
and that the dynamic of the model satisfies an additional technical condition 
which we shall explain in more detail later. 

In the {\em second section}, we give a short introduction
in the framework of algebraic quantum field theory in which   
a $1+1$ dimensional quantum field theory 
is described by a prescription which assigns to each bounded 
region $\2O\subset\7R^2$
a C*-algebra $\6A(\2O)$. The elements in $\6A(\2O)$ represent 
physical operations which are localized in $\2O$. 
This prescription has to satisfy a list of axioms
which are motivated by physical principles.

For our purpose it is convenient to work with 
the time slice formulation of a quantum field theory. 
We fix a {\em space-like plane} 
$\Sgm\subset\7R^2$ and consider a prescription which assigns 
to each bounded subset $\2I\subset \Sgm$ a C*-algebra $\6M(\2I)$.
The elements in $\6M(\2I)$ represent boundary conditions for 
physical operations at time $t=0$. We may interpret the algebras $\6M(\2I)$ as 
{\em Cauchy data}.
For our analysis, it is sufficient to consider the algebras of the 
massive free scalar-field at time $t=0$. They are given by
$$
\6M(\2I):=\{ e^{i\phi_0(f_1)+i\pi_0(f_2)} :\supp(f_j)\subset \2I\subset\Sgm\}''
$$
where $\phi_0$ is the free time zero-field, represented on 
Fock-space, and $\pi_0$ its canonical 
conjugate momentum. The double-prime $''$ denotes the bicommutant 
with respect to the algebra of bounded operators on Fock-space.
We denote by $C^*(\6M)$ is the C*-algebra which is generated by 
all algebras $\6M(\2I)$. The space-like translations, i.e. translations in
$\Sgm\cong\7R$, act as an automorphism-group $\{\al_\1x:\1x\in\7R\}$ 
on $C^*(\6M)$, where $\al_\1x$ maps $\6M(\2I)$ onto $\6M(\2I+\1x)$.

To describe the time development of a physical system, we consider a 
special class of one-parameter automorphism groups $\{\al_t:t\in\7R\}$ which 
are called {\em dynamics}.
Motivated by physical principles, 
they should satisfy the following list of axioms:
\begin{description}
\item[{\it (1)}]
The automorphisms $\al_t$ commute with the spatial translations $\al_\1x$.
\item[{\it (2)}]
The propagation speed, which is induced by 
the automorphism group $\{\al_t:t\in\7R\}$,  
is not faster than the speed of light, i.e. if an operator $a$
is localized in the open interval $(\1x,\1y)$, then the operator  
$\al_t(a)$ localized in $(\1x-|t|,\1y+|t|)$. 
\end{description}

There are familiar examples of dynamics, namely the dynamic of the 
massive free scalar field and the interacting 
dynamics of the $P(\phi)_2$-models \cite{GlJa1}.

We close the second section, discussing the connection between the 
time slice formulation of a quantum field theory 
and its corresponding formulation in two-dimensional Minkowski space. 

In the {\em third section}, 
we introduce a class of states which are of interest for our 
subsequent analysis. 
A {\em state} is described by a normed positive linear functional $\om$
on the C*-algebra $C^*(\6M)$. For an operator $a\in\6M(\2I)$, the 
value $\om(a)$ is the {\em expectation value} of 
a physical operation $a$ in the state $\om$. 
Since we want to discuss vacuum states and 
states with particle-like properties, we select the 
class of states which satisfy the {\em Borchers criterion} 
(positivity of the energy). A state $\om$ fulfills the 
{\em Borchers criterion} if the conditions, listed below, are 
satisfied.
\begin{description}
\item[{\it (1)}] There exists a unitary strongly continuous representation
of the translation group $U:(t,\1x)\mapsto U(t,\1x)$ on the 
GNS-Hilbert space $\2H$ of $\om$ 
which implements $\al_{(t,\1x)}=\al_t\circ \al_\1x$ 
in the GNS-representation $\pi$ of $\om$, i.e.
$$
\pi\circ\al_{(t,\1x)}=\Ad(U(t,\1x))\circ\pi \ \ \ .
$$
This condition can physically be interpreted as the fact that  
the outcome of an experiment will not change if we prepare the 
same state in a translated laboratory.   
\item[{\it (2)}] The spectrum of the generator of 
$U(t,\1x)$ is contained in the closed forward light cone.
The physical interpretation of this {\em spectrum condition}
is the requirement that the energy has to be positive. This condition 
describes the stability of a physical system.  
\end{description}
In addition to the Borchers criterion, a {\em vacuum state} $\om_0$ 
is translationally invariance, i.e.:
$$
\om_0\circ\al_{(t,\1x)}=\om_0
$$
We have to mention, that in our context the definitions given above
{\em depend on the dynamic} of the specific model. 

It is shown by Glimm and Jaffe \cite{GlJa1} 
that for each dynamic of a $P(\phi)_2$-model, 
there exists a {\em vacuum state} $\om$. In some cases there are 
more than one vacuum state with respect to the same dynamic, for example 
there are two different vacua with respect to the $\phi^4_2$-dynamic
\cite{Froh1,GlJa1}.

A mathematical definition of  kink states and the main result of this paper 
are given in the {\em 4th section}.
A {\em kink state} $\om$ which interpolates vacuum states
$\om_1,\om_2$ is characterized by the following properties:
\begin{description}
\item[{\it Particle-like properties:}] 
We require that a kink state fulfills the Borchers criterion. This property
guarantees that one has the possibility to "move" a kink like a particle.
If the lower bound of the spectrum of $U(x)$ is 
an isolated mass shell, then a kink state "behaves" completely like 
a particle. 
\item[{\it The interpolation property:}]
A pair of vacuum states $\om_1,\om_2$ is interpolated by a 
kink state $\om$ if there is a bounded region $\2I\subset \Sgm$, 
such that \newline
$\om(a)=\om_1(a)$, if $a$ is localized in the 
left (space-like) complement of $\2I$, and $\om(a)=\om_2(a)$,
if $a$ is localized in the right (space-like) complement of $\2I$. 
In other words, the state $\om$ "looks like" the 
vacuum $\om_1$ at plus space-like infinity and it "looks like" 
the vacuum $\om_2$ at minus space-like infinity.
\end{description}

We are now prepared to formulate the main result.
\begin{description}
\item[The main Result:]
Let $(\om_1,\om_2)$ be a pair of two inequivalent 
vacuum states with respect to 
a dynamic of a $P(\phi)_2$-model, then there exists an interpolating
kink state $\om$.
\end{description}

The construction of an interpolating kink state is based on a 
simple physical idea. 
Let us consider a physical system in one spatial dimension,
represented by a net of observables (v.Neumann algebras) 
$\2I\mapsto\6M(\2I)$. As described above, 
we denote by $C^*(\6M)$ the C*-algebra which is generated by all local 
algebras $\6M(\2I)$.  

Let us suppose that there is an partition wall, represented by 
a bounded interval $\2I=(\1x,\1y)$, such that it  
splits our system into two infinitely
extended laboratories, namely the laboratory on the left side of the wall, i.e.
the region 
$\2I_{LL}:=(-\infty,\1x)$, and the laboratory on the right side of the wall, 
i.e. the region $\2I_{RR}=(\1y,\infty)$.  

The physical operations which take place in 
the laboratory on the left side of the wall are represented by the 
C*-algebra $C^*(\6M,\2I_{LL})$ which is 
generated by all local algebras $\2M(\2I)$, $\2I\subset \2I_{LL}$. 
Analogously we 
consider the physical operations, represented by the algebra 
$C^*(\6M,\2I_{RR})$, on the right side of the wall. 

The property of the wall to separate the left from the right 
laboratory can be mathematically 
formulated by the requirement 
that the C*-algebra which is generated 
by  $C^*(\6M,\2I_{LL})$ and $C^*(\6M,\2I_{RR})$ is isomorph to their 
C*-tensor product, i.e.:
\beq\label{iso}
C^*(\6M,\2I_{LL}\cup \2I_{RR}))
\cong C^*(\6M,\2I_{LL})\otimes C^*(\6M,\2I_{RR})
\eeq
This means that observations which take place in the left 
laboratory are {\em statistically independent} from those in the 
right one. See also \cite{Roo,Sak} for these notions. 

Let us suppose that our physical system possesses at least two inequivalent 
vacuum states $\om_1$ and $\om_2$. Since the partition wall, 
which plays the role of the kink region,
has the separation property, described above,  
the vacuum states $\om_1$ and $\om_2$ can independently be 
prepared in the laboratory on the left side
and in the laboratory on the right side respectively.

Let us give a mathematical description of this scenario. 
By using the 
isomorphy, which is described by equation (\ref{iso}), 
we conclude that the prescription
$$
ab\mapsto \om_1(a)\om_2(b) \ \ 
\mbox{$a\in C^*(\6M,\2I_{LL})$ and $b\in C^*(\6M,\2I_{RR})$}
$$
defines a state $\om$ on the C*-algebra 
$C^*(\6M,\2I_{LL}\cup \2I_{RR})$. By the Hahn-Banach theorem,
we know that there exists an extension $\hat\om$ of the state $\om$ to the 
algebra $C^*(\6M)$. 

The state $\hat\om$ interpolates the vacua 
$\om_1$ and $\om_2$ correctly, but
for an explicit construction 
of an extension of $\om$ which satisfies the {\em Borchers criterion}, 
some technical difficulties have to be overcome. 

To solve these problems, we use a technical trick (compare also 
\cite[chapter 5]{Froh1}), namely we couple 
two copies of our physical system, i.e. we consider the net\newline
$\2I\mapsto \6F_2(\2I):=\6M(\2I) \olt \6M(\2I)$ (W*-tensor product). 

The map $\al_F$ which is given by interchanging the
tensor factors, 
$$
\al_F:a_1\otimes a_2\mapsto a_2\otimes a_1
$$
is called the {\em flip automorphism}.
We interpret the algebra $\6F_2(\2I)$ as a field algebra with an internal 
$\7Z_2$-symmetry. For an unbounded region $\2J\subset\Sgm$, 
let us denote by $\6F_2(\2J)$ the 
{\em v.Neumann algebra}
which is generated by all algebras $\6F_2(\2I)$ with $\2I\subset\2J$.

We shall show in the appendix that for each bounded interval 
$\2I=(\1x,\1y)$ ($\2I_{RR}:=(\1y,\infty)$ and $\2I_R:=(\1x,\infty)$) 
the inclusion 
\beqa
\6F_2(\2I_{RR})\subset \6F_2(\2I_R)
\eeqa 
is split.
By using the universal localizing map with respect to this split inclusion,
a unitary operator $\te_\2I$ can be constructed, such that 
$\te_\2I$ implements the flip $\al_F$ on $\6F_2(\2I_{RR})$ and 
commutes with each element in $\6F_2(\2I_{LL})$ \cite{AntLo,Schl4,Mue1}. 
The adjoint action
of $\te_{\2I}$ induces an automorphism $\beta^{\2I}$ of 
$C^*(\6F_2)$ which has the following properties:  
\begin{description}
\item[{\it (1)}]
The automorphism $\beta^{\2I}$ is an involution, i.e. 
$\beta^{\2I}\circ\beta^{\2I}=\id$.
\item[{\it (2)}]
For $a\in C^*(\6F_2,\2I_{LL})$ and $b\in C^*(\6F_2,\2I_{RR})$ 
one has:
$$
\beta^{\2I}(a)=\al_F(a) \ \ \mbox{  and  } \ \ 
\beta^{\2I}(b)=b
$$
\end{description}

For each bounded interval $\2I$, the automorphism $\beta^\2I$ can be used to 
construct an extension $\hat\om$ of the state $\om$ to the algebra 
$C^*(\6M)$, namely:
$$
\hat\om:=\om_1\otimes\om_2\circ\beta^\2I|_{C^*(\6M)\otimes\7C\11}
$$
We shall show that the state $\hat\om$ satisfies the Borchers criterion if 
the automorphism 
\beqa
\al_{(t,\1x)}\circ\beta^\2I\circ\al_{-(t,\1x)}\circ\beta^\2I
\eeqa
of $C^*(\6F_2)$ is inner, 
i.e. it is given by the adjoint action of a local operator 
$\gam(t,\1x)\in C^*(\6F_2)$.
Indeed, for this case, the translation group 
is implemented in the GNS-representation 
of $\hat\om$ by the representation
\beqa
(t,\1x)\mapsto 
U(t,\1x)=U_1(t,\1x)\otimes U_2(t,\1x)\pi_1\otimes\pi_2(\gam(-t,-\1x))
\eeqa
where $U_1(t,\1x)$ and $U_2(t,\1x)$ implement the translations 
in the corresponding vacuum representations $\pi_1$ and $\pi_2$.
The spectrum condition can be proven by using the additivity of 
energy-momentum spectrum, as described in \cite{Schl4}. 

We shall show in {\em section 5} that  
the automorphism 
\beqa
\beta^\2I\circ\al_{(t,\1x)}\circ\beta^\2I\circ\al_{-(t,\1x)}
\eeqa
is inner in $C^*(\6F_2)$ if the dynamic $\al$ of the 
model under consideration can be extended to the (non-local) net \newline
$\2I\mapsto\hat\6F_2(\2I):=\6F_2(\2I)\vee\{\te_\2I\}$
(compare also \cite{Mue1}).
Once we have established this result, we conclude that 
$\hat\om$ satisfies the Borchers criterion.

These aspects are discussed for a slightly 
more general situation. We consider 
the net which is given by the 
$N$-fold W*-tensor product \newline
$\2I\mapsto\6F_N(\2I):=\6M(\2I)^{\olt N}$. 
The permutation group $S_N$ acts on it as an internal symmetry group 
of automorphisms $\{\al_\sgm;\sgm\in S_N\}$.

This generalization can be used to construct {\em multi-kink states}
in a very simple way. Given a permutation $\sgm\in S_N$ and 
a bounded interval $\2I$. Using the universal localizing map, we  
construct an automorphism $\al_\sgm^\2I$ which acts on observables, 
localized in $\2I_{LL}$, trivially 
and on observables, 
localized in $\2I_{RR}$,
as the automorphism $\al_\sgm$.

Let us consider a family of vacuum states 
$(\om_1,\cdots,\om_N)$ and intervals \newline 
$\2I_1,\cdots,\2I_N$. Then the 
state 
$$
\om:=\om_1\otimes\cdots\otimes\om_N
\circ\al_{s_N}^{\2I_N}\cdots\al_{s_1}^{\2I_1}|_{C^*(\6M)\otimes\7C\11}
$$
can be interpreted as a multi-kink-state. Here $s_j$ denotes the 
transposition of $j$ and $j+1$.

Indeed, 
if we consider an 
observable $a\in\6M(\2I)$ with $\2I>\2I_1\cup\cdots\cup\2I_N$ we obtain
\begin{eqnarray*}
\om(a)=\om_1\otimes\cdots\otimes\om_N
\circ\al_{s_N}^{\2I_N}\cdots\al_{s_1}^{\2I_1}(a\otimes\11)
\vs\vs
=\om_1\otimes\cdots\otimes\om_N
\circ\al_{s_N\cdots s_1}(a\otimes\11)
=\om_N(a)
\end{eqnarray*}
Analogously we obtain for $a'\in\6M(\2I')$ with 
$\2I'<\2I_1\cup\cdots\cup\2I_N$
$$
\om(a')=\om_1(a')
$$ 
since the automorphisms $\al_{s_j}^{\2I_j}$ act trivially on $\6M(\2I')$.

Finally, to prove the main result of 
our paper, we shall show in {\em section 6} that 
each dynamic of a 
$P(\phi)_2$-Model is extendible.

We close our paper with {\em section 7}, where we give 
a summary of the main results and discussing some work 
in progress.

%%%%%%%%%%%%%%%%%%%%%%%%%%%%%%%%%%%%%%%%%%%%%%%%%%%%%%%%%%%%%%%%%%%%%%%%%%%
\section{Preliminaries}
%%%%%%%%%%%%%%%%%%%%%%%%%%%%%%%%%%%%%%%%%%%%%%%%%%%%%%%%%%%%%%%%%%%%%%%%%%%

%%%%%%%%%%%%%%%%%%%%%%%%%%%%%%%%%%%%%%%%%%%%%%%%%%%%%%%%%%%%%%%%%%%%%%%%%%
\paragraph*{The Framework of Algebraic Quantum Field Theory:}
%%%%%%%%%%%%%%%%%%%%%%%%%%%%%%%%%%%%%%%%%%%%%%%%%%%%%%%%%%%%%%%%%%%%%%%%%%

Let us consider a quantum field theory in two dimensions which is described 
by a translationally covariant Haag-Kastler net. We briefly discuss 
the axioms of such a net.

A $1+1$ dimensional quantum field theory 
is given by a prescription which assigns to each region $\2O\subset\7R^2$
a C*-algebra $\6A(\2O)$ and the elements in $\6A(\2O)$ represent 
physical operations which are localized in $\2O$. 
This prescription has to satisfy a list of axioms
which are motivated by physical principles.
\begin{description}
\item[{\it (1)}]
A physical operation
which is localized in a region $\2O$ should also localized in each 
region which contains $\2O$. Therefore, we require that   
if a region 
$\2O_1$ is contained in a lager region $\2O$, then the algebra $\6A(\2O_1)$
is a sub-algebra of $\6A(\2O)$. 
\item[{\it (2)}]
Two local operations which take place 
in space-like separated regions should not influence each other.
Hence the {\em principle of locality} is formulated as follows:
If a region 
$\2O_1$ is space-like separated from a region $\2O$, 
then the elements of $\6A(\2O_1)$ commute with those of $\6A(\2O)$. 
\item[{\it (3)}]
Each operation which is localized in $\2O$ should have an 
equivalent counterpart which is localized in a translated region
$\2O+x$. The {\em principle of translation covariance} is described by 
the existence of a 
two-parameter automorphism group $\{\al_x;x\in\7R^2\}$ which 
acts on the C*-algebra $\6A$, 
generated by all local algebras $\6A(\2O)$, such that 
$\al_x$ maps $\6A(\2O)$ onto $\6A(\2O+x)$.
\end{description}
A prescription $\2O\to\6A(\2O)$ of this type is called a {\em translationally
covariant Haag-Kastler net}.

%%%%%%%%%%%%%%%%%%%%%%%%%%%%%%%%%%%%%%%%%%%%%%%%%%%%%%%%%%%%%%%%%%%%%%%%%%
\paragraph*
{Cauchy Data and Dynamics of a Quantum Field Theory:}
%%%%%%%%%%%%%%%%%%%%%%%%%%%%%%%%%%%%%%%%%%%%%%%%%%%%%%%%%%%%%%%%%%%%%%%%%%

For our purpose, it is convenient to work with the time slice formulation 
of a quantum field theory. Let us choose a space-like plain 
$\Sgm\subset\7R^2$. The time slice-formulation has two main aspects.
Firstly, the {\em Cauchy data} with respect to $\Sgm$ which describes the 
boundary conditions at time $t=0$. Second, the {\em dynamic} which 
describes the time evolution of the quantum fields. 

The {\em Cauchy data} of a quantum field theory are given by a 
net of v.Neumann-algebras
$$
\6M:=\{\6M(\2I)\subset\6B(\2H_0);
\mbox{ $\2I$ is open and bounded interval in $\Sgm$} \}
$$
represented on a Hilbert-space $\2H_0$. This net has to 
satisfy the following conditions:
\bdes
\itno 1
The net is satisfies isotony, i.e. if  
$\2I_1\subset\2I_2$, then $\6M(\2I_1)\subset\6M(\2I_2)$.
\itno 2
The net is local, i.e. if 
$\2I_1\cap\2I_2=\emptyset$, then $\6M(\2I_1)\subset\6M(\2I_2)'$.
\itno 3
There exists a unitary and strongly continuous representation
$$
U:\1x\in \7R \mapsto U(\1x)\in\2U(\2H_0)
$$
of the 
spatial translations in $\Sgm\cong\7R$, such that 
$\al_{\1x}:=\Ad(U(\1x))$ maps $\6M(\2I)$ onto $\6M(\2I+\1x)$.
\edes

\paragraph*{\it Notation:}
Let us give a few comments on the notation to be used.
Given a net $\6N:\2I\mapsto\6N(\2I)$ of W*-algebras.
In the sequel, we denote the 
C*-inductive limit of the net $\6N$ by $C^*(\6N)$.
The corresponding C*- and W*-algebras, which belong to an 
unbounded region $\2J\subset\Sgm$, are denoted by
$$
C^*(\6N,\2J):=\ol{\bigcup_{\2I\subset\2J}\6N(\2I)}^{||\cdot||}
\ \ \mbox{ and } \ \ 
\6N(\2J):=\bigvee_{\2I\subset\2J}\6N(\2I) \ \ \ \mbox{ respectively.}
$$
Furthermore,we write $\aut(\6N)$, for the 
group of *-automorphisms of $C^*(\6N)$.

\paragraph*{\it Remark:}
In general, the W*-algebra  $\6N(\2J)$ is {\em not contained} in 
the C*-inductive limit $C^*(\6N)$, since $C^*(\6N)$ is only generated by 
algebras with respect to bounded intervals. 
\paragraph*{}

As mentioned in the introduction, we introduce the notion of of dynamics.
%%%%%%%%%%%%%%%%%%%%%%%%%%%%%%%%%%%%%%%%%%%%%%%%%%%%%%%%%%%%%%%%%%%%%%%%%%%%%
\bdef
\label{def1}
A one-parameter group of automorphisms \newline
$\al=\{\al_t\in\aut(\6M);t\in\7R\}$
is called a {\em dynamic} of the net $\6M$ if
\begin{description}
\item[{\it (1)}] 
the automorphism group $\al$ has {propagation speed} $\ps(\al)\leq 1$, where 
$\ps(\al)$ is defined as follows:
$$
\ps(\al):=
\inf\{\beta'|\al_t\6M(\1x,\1y)\subset \6M(\1x-\beta'|t|,\1y+\beta'|t|);
\forall t,\1x,\1y\}
$$
\item[{\it (2)}] 
The automorphisms $\{\al_t\in\aut(\6M);t\in\7R\}$ 
commute with the automorphism-group 
of spatial translations $\{\al_\1x\in\aut(\6M);\1x\in\7R\}$, i.e.:
$$
\al_t\circ\al_\1x=\al_\1x\circ\al_t \ \ \ ; \ \ \ \forall \1x,t
$$
\end{description}
The set of all dynamics of $\6M$ is denoted by $\dyn(\6M)$.
\eef  
%%%%%%%%%%%%%%%%%%%%%%%%%%%%%%%%%%%%%%%%%%%%%%%%%%%%%%%%%%%%%%%%%%%%%%%%%%%%
Here we write $\6M(\1x,\1y)$ for the algebra which belongs to the 
interval \newline $\2I=(\1x,\1y)$.

At this point we should mention that it is possible to 
choose for different theories the same net of Cauchy data. 
In case of $P(\Phi)_2$-models, the Cauchy data are given by 
the time zero-algebras of the free massive scalar field.

To distinguish different theories, we have to compare different dynamics.
For this purpose, 
we shall construct 
a {\em universal Haag-Kastler net} 
with respect to the net $\6M$ of 
Cauchy data in the next paragraph.

%%%%%%%%%%%%%%%%%%%%%%%%%%%%%%%%%%%%%%%%%%%%%%%%%%%%%%%%%%%%%%%%%%%%%%%%%%%%
\paragraph*{Haag-Kastler nets for Cauchy Data:}
%%%%%%%%%%%%%%%%%%%%%%%%%%%%%%%%%%%%%%%%%%%%%%%%%%%%%%%%%%%%%%%%%%%%%%%%%%%%

We denote by $U(\6M)$ the group of unitary operators in $C^*(\6M)$.
Let $\6G(\7R,\6M)$ be the group which is generated by 
the set 
$$
\{ (t,u)|\mbox{ $t\in\7R$ and $u\in U(\6M)$ } \}
$$
modulo the following relations:
\bdes
\itno 1
For each $u_1,u_2\in U(\6M)$ and for each $t_1,t_2,t\in\7R$, we require: 
$$
(t,u_1)(t,u_2)=(t,u_1u_2) \mbox{ and } (t,\11)=\11
$$
\itno 2
For $u_1\in\6M(\2I_1)$ and $u_2\in\6M(\2I_2)$ with 
$\2I_1\subset\2I_2+[-|t|,|t|])$
we require for each $t_1\in\7R$:
$$
(t_1+t,u_1)(t_1,u_2)=(t_1,u_2)(t_1+t,u_1) 
$$
\edes
We conclude from relation {\it (1)} that $(t,u)$ is the inverse 
of $(t,u^*)$. Furthermore, a localization 
region in $\7R\times\Sgm$ can be assigned
to each element in $\6G(\7R,\6M)$. 

An element of the form 
$$
v=(t_1,u_1)\cdots(t_n,u_n)
$$
is localized in $\2O\subset\7R\times\Sgm$ if the following holds:
\bdes
\item[]
There exists a region $\2I\subset\Sgm$, such that  
$\{t_1,\cdots ,t_n\}\times\2I\subset\2O$ and \newline
$u_1,\cdots, u_n\in\6M(\2I)$.
\edes
The subgroup of $\6G(\7R,\6M)$ which is generated by elements which are  
localized in the double cone $\2O$, is denoted by $\6G(\2O)$.
 
We easily observe that relation {\it (2)} implies that group elements commute 
if they are localized in space-like separated regions.  

The translation group in $\7R^2$ is naturally represented by  
group-automorphisms of $\6G(\7R,\6M)$. They are defined 
by the prescription  
$$
\beta_{(t,\8x)}(t_1,u):=(t+t_1,\al_\8xu) \ \ \ .
$$
Thus the subgroup $\6G(\2O)$ is mapped onto $\6G(\2O+(t,\8x))$
by $\beta_{(t,\8x)}$.

To construct the universal Haag-Kastler net, we build the group C*-algebra
$\6B(\2O)$ with respect to $\6G(\2O)$. For convenience, we briefly describe
the construction of $\6B(\2O)$.

In the first step we build the *-algebra $\6B_0(\2O)$ which is generated 
by all complex valued functions $a$ on $\6G(\2O)$, such that 
\beqa
a(u)=0 \ \ \mbox{ for almost each $u\in\6G(\2O)$ .} 
\eeqa
We write such a function symbolically as a formal sum, i.e.
\beqa
a=\sum_u a(u)\ u
\eeqa
The product and the *-relation is given as follows:
\beqa
ab= \sum_u a(u) \ u \cdot \sum_{u'} b(u') \ u' =
\sum_{u'}\bl(\sum_u a(u)b(u^{-1}u')\br) u'
\vs\vs
a^*=\sum_u \bar a(u^{-1}) \ u
\eeqa

It is well known, that the algebra $\6B_0(\2O)$ has a 
C*-norm which is given by
\beqa
||a||:=\sup_\pi||\pi(a)||_\pi
\eeqa
where the supremum is taken over each Hilbert-space representation $\pi$ 
of $\6B_0(\2O)$. Finally, we define $\6B(\2O)$ as the closure of 
$\6B_0(\2O)$ with respect to the norm above.

The C*-algebra which is generated by all local algebras $\6B(\2O)$ 
is denoted by $C^*(\6B)$. 
By construction, the group isomorphisms $\beta_{(t,\8x)}$ 
induce a representation of the translation group by automorphisms 
of $C^*(\6B)$. 

\paragraph*{\it Observation:}
The net of C*-algebras
$$
\6B:=\{\6B(\2O)|\2O \ \mbox{ is a bounded double cone in $\7R^2$ }\}
$$ 
is a translationally covariant Haag-Kastler net. 
\paragraph*{}

The universal properties of the net $\6B$
are stated in the following Proposition:

%%%%%%%%%%%%%%%%%%%%%%%%%%%%%%%%%%%%%%%%%%%%%%%%%%%%%%%%%%%%%%%%%%%%%%%%%%%%%
\bpro\label{pro0}
Each dynamic $\al\in\dyn(\6M)$ induces a C*-homomorphism 
$$
\iota_\al:C^*(\6B)\to C^*(\6M)
$$
such that
$$
\iota_\al\circ\beta_{(t,\8x)}=\al_{(t,\8x)}\circ\iota_\al \ \ \mbox{ ,}
$$
for each $(t,\8x)\in\7R^2$.
In particular,   
$$
\2O\mapsto\6A_\al(\2O):=\iota_\al(\6B(\2O))'' 
$$
is a translationally covariant Haag-Kastler net.
\epro
%%%%%%%%%%%%%%%%%%%%%%%%%%%%%%%%%%%%%%%%%%%%%%%%%%%%%%%%%%%%%%%%%%%%%%%%%%%%
\bpr
Given a dynamic $\al$ of $\6M$.
We conclude from $\ps(\al)\leq 1$ that the prescription 
$$
(t,u)\mapsto \al_tu  
$$
defines a C*-homomorphism  
$$
\iota_\al:C^*(\6B)\to C^*(\6M)\ \ .
$$
In particular, $\iota_\al$ is a representation of $C^*(\6B)$
on the Hilbert space $\2H_0$. 
This statement can be obtained by using the relations, listed below.
\bdes
\itno a
$$
\iota_\al((t,u_1)(t,u_2))=\al_tu_1\al_tu_2=\al_t(u_1u_2)=
\iota_\al(t,u_1u_2)
$$
\itno b
If $(t_1,u_1)$ and $(t_1+t,u_2)$ are localized in space-like separated 
regions, then we obtain from $\ps(\al)\leq 1$:
$$
[\iota_\al(t_1,u_1),\iota_\al(t_1+t,u_2)]=
\al_{t_1}[u_1,\al_tu_2]=0
$$
\itno c
$$
\iota_\al(\beta_{(t,\8x)}(t_1,u))=\iota_\al(t+t_1,\al_{\8x}u)=
\al_{(t,\8x)}\al_{t_1}u
$$
\edes
\epr
%%%%%%%%%%%%%%%%%%%%%%%%%%%%%%%%%%%%%%%%%%%%%%%%%%%%%%%%%%%%%%%%%%%%%%%%%%

In general we expect that for a given dynamic $\al$ the representation 
$\iota_\al$ is not faithful. Hence each dynamic defines a two-sided 
ideal 
$$
J(\al):=\iota_\al^{-1}(0)\in C^*(\6B)
$$
in $C^*(\6B)$ which we call the {\em dynamical ideal} 
with respect to $\al$ and the quotient C*-algebras 
\beqa
\6B(\2O)/J(\al)\cong\6A_\al(\2O)
\eeqa
may depend on the dynamic $\al$. Indeed, if $\2O$ is a double cone 
whose base is {\em not contained} in $\Sgm$, then 
for different dynamics $\al_1,\al_2$ the algebras $\6A_{\al_1}(\2O)$ and
$\6A_{\al_2}(\2O)$ are different. On the other hand, if the base 
of $\2O$ is contained in $\Sgm$, then we conclude from 
the fact that the dynamic $\al$ has finite propagation speed and 
from Proposition \ref{pro0}:
%%%%%%%%%%%%%%%%%%%%%%%%%%%%%%%%%%%%%%%%%%%%%%%%%%%%%%%%%%%%%%%%%%%%%%%%%%%
\bcor
If $\2I\subset\Sgm$ is the base of the double cone $\2O$, then the algebra
$\6A_\al(\2O)$ is independent of $\al$.
In particular, the C*-algebra 
$$
C^*(\6M)
=\ol{\bigcup_\2I \6M(\2I)}^{||\cdot||} 
=\ol{\bigcup_\2O \6A_\al(\2O)}^{||\cdot||} 
$$
is the C*-inductive limit of the net $\6A_\al$.
\ecor
%%%%%%%%%%%%%%%%%%%%%%%%%%%%%%%%%%%%%%%%%%%%%%%%%%%%%%%%%%%%%%%%%%%%%%%%%%%
From the discussion above, we see that two dynamics with the same dynamical 
ideal induces the same quantum field theory. 

%%%%%%%%%%%%%%%%%%%%%%%%%%%%%%%%%%%%%%%%%%%%%%%%%%%%%%%%%%%%%%%%%%%%%%%%%%%%
\paragraph*{The Massive Free Scalar Field:}
%%%%%%%%%%%%%%%%%%%%%%%%%%%%%%%%%%%%%%%%%%%%%%%%%%%%%%%%%%%%%%%%%%%%%%%%%%

As mentioned above, the Cauchy data for the $P(\phi)_2$-models
are given by the time zero-algebras of the massive free scalar field.
For our purpose, let us briefly describe the time slice formulation of the 
massive free scalar field in one spatial dimension.

Let us denote by $\2H_0$ the symmetrized Fock space over 
$L_2(\7R)$, i.e.
$$
\2H_0=\bigoplus_{n=0}^\infty s_n (L_2(\7R)^{\otimes n})
$$
where $s_n$ denotes the symmetrization operator.
As usual, we consider annihilation and creation operators, where the 
creation operator is given by 
$$
a^*(f)\psi = \sum_{n=0}^\infty n^{1/2}s_n(\hat f\otimes \Pi_{n-1}\psi) \ \ \ .
$$
$\Pi_n$ denotes the canonical projection from $\2H_0$ onto 
$\2H_{0,n}:=s_n(L_2(\7R)^{\otimes n})$ and $\hat f$ 
the Fourier transform of $f$. The operator $a^*(f)$ has an 
adjoint\newline
$a(f):=(a^*(f))^*$ which is the annihilation operator.

There is a unitary and strongly continuous representation of the 
spatial translation group on $\2H_0$, which is given by
$$ 
\Pi_n(U(\1x)\psi)(\1k_1\cdots\1k_n)= \exp\biggl( i\1x\sum_{i=1}^n \1k_i\biggr)
\Pi_n\psi(\1k_1\cdots\1k_n) \ \ \ ,
$$
together with a unique vector $\Om_0\in \2H_0$ which is invariant under the 
translation group $\{ U(\1x); \1x\in\7R\}$, namely 
$$
\Om_0=(1,0,0,0,\cdots) \ \ \ .
$$
 
The massive free Bose field at time $t=0$ is an  
operator valued distribution $B$ on $K=S_\7R(\7R)\oplus S_\7R(\7R)$. 
For a function $f=f_1\oplus f_2\in K$ the operator $B(f)$ is 
given by
$$
B(f):=
{1\over 2}\biggl (a^*(\mu^{-1/2}f_1)+a(\mu^{-1/2}f_1)\biggr)
+{1\over 2i}\biggl (a^*(\mu^{1/2}f_2)-a(\mu^{1/2}f_2) \biggr)
$$
where $\mu^\tau$ is the 
pseudo differential operator which is given by kernel
\beq
\mu^\tau(\1x-\1y)
:=\int \8d\1p \ (\1p^2+m^2)^{\tau/2} \ e^{i\1p(\1x-\1y)} \ \ \ .
\eeq
It is well known that $B(f)$ is an essentially self adjoint 
operator. 

\paragraph*{\it Notation:}
Given a region $G\subset\7R$.
We denote by $\6M(G)$ the v.Neumann algebra which is given by
$$
\6M(G):=\{ w(f):=e^{iB(f)}: \supp(f)\subset G \}'' \ \ \ ,
$$ 
where $''$ denotes the 
bicommutant in $\6B(\2H_0)$. 
\paragraph*{}

Hence we obtain a net of Cauchy data:
$$
\6M:=\{\6M(\2I)|\mbox{ $\2I$ is open and bounded interval in $\7R$}  \}
$$
The algebras with respect to half-lines, for example $G=(\1x,\infty)$, 
are called {\em wedge algebras}. They
play an important role for the construction of kink-states. 

\paragraph*{\it Notation:} 
Let us consider a bounded interval $\2I=(\1x,\1y)$. We 
define the following four regions notation with respect to $\2I$: 
$$
\begin{array}{l}
\2I_{LL}:=(-\infty,\1x) \ \ \mbox{ , } \ \ \2I_L:=(-\infty,\1y) \ \ \mbox{ , }
\vs
\2I_R:=(\1x,\infty) \ \ \mbox{ and } \ \  \2I_{RR}:=(\1y,\infty) \ \ \ .
\end{array}
$$
\paragraph*{}

An important
property, which we shall use later, is given in the 
proposition below.

%%%%%%%%%%%%%%%%%%%%%%%%%%%%%%%%%%%%%%%%%%%%%%%%%%%%%%%%%%%%%%%%%%%%%%%%%%
\bpro\label{pro1}
Given a nonempty and bounded interval $\2I$. Then the inclusion 
$$
\6M(\2I_{RR})\subset \6M(\2I_R)
$$
is standard split. 
\epro
%%%%%%%%%%%%%%%%%%%%%%%%%%%%%%%%%%%%%%%%%%%%%%%%%%%%%%%%%%%%%%%%%%%%%%%%%%
\bpr 
The proof of the statement can be found in the appendix.  
Here the methods of \cite{Bu1} are used. Compare also the results of 
\cite{AntFre,AntLo}.
\epr
%%%%%%%%%%%%%%%%%%%%%%%%%%%%%%%%%%%%%%%%%%%%%%%%%%%%%%%%%%%%%%%%%%%%%%%%% 

Since the inclusion which is given above is standard split, 
there exists a unitary operator 
$$
w_{\2I}:\2H_0\to \2H_0\otimes \2H_0  
$$
such that for $a\in \6M(\2I_{LL})$ and $b\in\6M(\2I_{RR})$ we have: 
$$
w_{\2I}(a\otimes b)w_{\2I}^*=ab
$$
Thus there is an interpolating type I factor $\2N\cong\6B(\2H_0)$, i.e.
$$
\6M(\2I_{RR})\subset\2N\subset \6M(\2I_R)
$$
which is given by 
$$
\2N:=w_{\2I}(\11\otimes \6B(\2H_0))w_{\2I}^* \ \ \ .
$$
Hence we obtain an embedding of $\6B(\2H_0)$ into the algebra 
$\6M(\2I_R)$:
$$
\Psi_{\2I}:F\in\6B(\2H_0)\mapsto 
w_{\2I}(\11\otimes F)w_{\2I}^*\in\6M(\1x,\infty)
$$
This embedding is called the {\em universal localizing map}.

%%%%%%%%%%%%%%%%%%%%%%%%%%%%%%%%%%%%%%%%%%%%%%%%%%%%%%%%%%%%%%%%%%%%%%%%%%
\section{States}
%%%%%%%%%%%%%%%%%%%%%%%%%%%%%%%%%%%%%%%%%%%%%%%%%%%%%%%%%%%%%%%%%%%%%%%%%
Let us consider the set $\6S$ of all {\em locally normal states}
on $C^*(\6M)$, i.e. for each state $\om\in\6S$ and for each bounded interval
$\2I$, the restriction
$$
\om|_{\6M(\2I)}
$$
is a normal state on $\6M(\2I)$. 

As mentioned in the introduction, we are interested in 
states with vacuum and particle-like properties, i.e. 
states which satisfies the {\em Borchers criterion} 
(See the Introduction for this notion).

\paragraph*{\it Notation:}
Given a dynamic $\al\in\dyn(\6M)$. We denote the corresponding 
set of all locally normal
states which satisfies the Borchers criterion 
by $\6S(\al)$ and analogously the set of all vacuum states by $\6S_0(\al)$.
Moreover, we write for the set 
of vacuum sectors 
\beq
\sec_0(\al):=\{[\om]|\om\in\6S_0(\al) \}
\eeq
where $[\om]$ denotes the unitary equivalence class of the 
the GNS-representation of $\om$. 
\paragraph*{}

In the next two paragraphs, we discuss some familiar 
examples of vacuum states.

%%%%%%%%%%%%%%%%%%%%%%%%%%%%%%%%%%%%%%%%%%%%%%%%%%%%%%%%%%%%%%%%%%%%%%%%%%%
\paragraph*{Free Vacuum States:}
%%%%%%%%%%%%%%%%%%%%%%%%%%%%%%%%%%%%%%%%%%%%%%%%%%%%%%%%%%%%%%%%%%%%%%%%%%%
The simplest example for a vacuum state is 
the free massive vacuum state $\om_0$ with respect to 
the free dynamic 
$$
\al_{0,t}(a)=e^{ih_0t}ae^{-ih_0t} 
$$
which is given by the free Hamiltonian 
$$
h_0=\int \8d\1p \ (\1p^2+m^2)^{1/2} \ a^*(\1p)a(\1p)
$$
As usual,  
$a(\1p)$ and $a^*(\1p)$ are the creation and annihilation forms on the 
Fock space $\2H_0$.
%%%%%%%%%%%%%%%%%%%%%%%%%%%%%%%%%%%%%%%%%%%%%%%%%%%%%%%%%%%%%%%%%%%%%%%%%%%%
\paragraph*{Vacuum States for Interacting Dynamics:}
%%%%%%%%%%%%%%%%%%%%%%%%%%%%%%%%%%%%%%%%%%%%%%%%%%%%%%%%%%%%%%%%%%%%%%%%%%%
Further examples for vacuum states are the vacua of 
the $P(\phi)_2$-models. 
The interacting part of the cutoff Hamiltonian is 
given by a Wick polynomial of the time zero field $\phi_0$, i.e.
$$
h_1(\2I)=h_1(\chi_\2I)=:P(\phi_0):(\chi_\2I)
$$
where $\chi_\2I$ is a test function with $\chi_\2I(\1x)=1$ for 
$\1x\in\2I$ and $\chi_\2I(\1y)=0$ 
on the complement of a slightly lager region $\hat\2I\supset\2I$.
It is well known that $h_1(\2I)$ is a self-adjoint operator, which 
has a joint core with the free Hamiltonian $h_0$, and is affiliated with
$\6M(\hat\2I)$.
The operator $h_1(\2I)$ induces a automorphism group 
$\al_\2I$ which is given by
$$
\al_{\2I,t}(a):=e^{ih_1(\2I)t}ae^{-ih_1(\2I)t} \ \ \ .
$$
Consider the inclusion of intervals $\2I_0\subset\2I_1\subset\2I_2$.
Then we have for each $a\in\6M(\2I_0)$:
$$
\al_{\2I_1,t}(a)=\al_{\2I_2,t}(a)
$$
Hence, there exists a one-parameter automorphism group \newline
$\{\al_{1,t}\in\aut(\6M);t\in\7R\}$ such that 
$\al_{1,t}$ acts on $a\in\6M(\2I)$ as follows:
$$
\al_{1,t}(a)=\al_{\2I,t}(a) \ \ \ ; \ \ \ \forall t\in\7R
$$
The automorphism group $\{\al_{1,t}\in\aut(\6M);t\in\7R\}$ is a 
dynamic of $\6M$ with zero propagation speed, i.e. $\ps(\al_1)=0$.

Since $h_1(\2I)$ has a joint core with the free Hamiltonian $h_0$,
we are able to define the Trotter product of the automorphism-groups
$\al_0$ and $\al_1$ which is given 
for each local operator $a\in\6M(\2I)$ by
\beqa
\al_t(a):=(\al_0\times\al_1)_t(a)=
s-\lim_{n\to\infty}(\al_{0,t/n}\circ\al_{1,t/n})^n(a)  \ \ \ .
\eeqa
The limit is taken in the strong operator topology.
Furthermore,the propagation speed is sub-additive with respect to the 
Trotter product \cite{GlJa1}, i.e.
$$
\ps(\al_0\times\al_1)\leq\ps(\al_0)+\ps(\al_1)
$$
and we conclude that 
$\al\in\dyn(\6M)$ is a dynamic of $\6M$. We call the dynamic $\al$ 
{\em interacting}. 

It is shown by Glimm and Jaffe \cite{GlJa1}
that there exist vacuum states 
$\om$ with respect to the interacting dynamic $\al$.
We have to mention, that there is {\em no} vector $\psi$ in 
Fock space $\2H_0$, such that the state
\beqa
a\mapsto\<\psi,a\psi\>
\eeqa
is a vacuum state with respect to an interacting dynamic $\al$, but there
is a net of vectors $(\Om_\Lam)$ in $\2H_0$ such that the limit
\beqa
\om=\lambda^*-\lim_\Lam \<\Om_\Lam,\cdot\Om_\Lam\>
\eeqa
is a 
vacuum state with respect to the dynamic $\al$.
The limit has to be taken in the local norm topology 
(here denoted by $\lambda^*$) on $C^*(\6M)^*$ which 
is induced by the following family of semi norms:
\beqa
\bl\{||\varphi||_\2I:=
\sup_{a\in\6M(\2I)}||a||^{-1}|\varphi(a)| \bm | \mbox{ $\2I$
is an open bounded interval }\br\}
\eeqa
Of course, the topology $\lambda^*$ is weaker than the ordinary norm topology 
and stronger than the weak*-topology.
In addition to that, the set of locally normal states $\6S$ 
is complete with respect to the topology $\lambda^*$.

%%%%%%%%%%%%%%%%%%%%%%%%%%%%%%%%%%%%%%%%%%%%%%%%%%%%%%%%%%%%%%%%%%%%%%%%%%%%%
\section{Interpolating Kink States}
%%%%%%%%%%%%%%%%%%%%%%%%%%%%%%%%%%%%%%%%%%%%%%%%%%%%%%%%%%%%%%%%%%%%%%%%%%%%%

In this section we give a mathematical definition of a kink state and 
we formulate the main result of our paper. 

\paragraph*{\it Notation:}
Let us write $(\2H,\pi,\Om),(\2H_j,\pi_j,\Om_j)$ for the GNS-triples of
the states $\om\in\6S(\6M)$ and $\om_j\in\6S_0(\6M);j=1,2$ respectively, 
unless we state something different.

%%%%%%%%%%%%%%%%%%%%%%%%%%%%%%%%%%%%%%%%%%%%%%%%%%%%%%%%%%%%%%%%%%%%%%%%%%%%
\paragraph*{Definition of Kink States:}
%%%%%%%%%%%%%%%%%%%%%%%%%%%%%%%%%%%%%%%%%%%%%%%%%%%%%%%%%%%%%%%%%%%%%%%%%%%
\bdef\label{def4}
Let $\al\in\dyn(\6M)$ be a dynamic of $\6M$.  
A state $\om$ of $\6M$ is called a {\em kink state}, 
interpolating vacuum states
$\om_1,\om_2\in\6S_0(\al)$ if 
\begin{description}
\item[{\it (a)}]
$\om$ satisfies the Borchers criterion
\item[{\it (b)}]
and there exists a bounded interval $\2I$, such that $\om$ 
fulfills the relations:
$$
\pi|_{C^*(\6M,\2I_{LL})}\cong\pi_1|_{C^*(\6M,\2I_{LL})} \ \ \mbox{ and } \ \ 
\pi|_{C^*(\6M,\2I_{RR})}\cong\pi_2|_{C^*(\6M,\2I_{RR})}    
$$
The symbol $\cong$ means unitarily equivalent.
\end{description}
The set of all kink states which interpolate $\om_1$ and $\om_2$ is denoted by
\newline
$\6S(\al|\om_1,\om_2)$.
\eef
%%%%%%%%%%%%%%%%%%%%%%%%%%%%%%%%%%%%%%%%%%%%%%%%%%%%%%%%%%%%%%%%%%%%%%%%%%%%

%%%%%%%%%%%%%%%%%%%%%%%%%%%%%%%%%%%%%%%%%%%%%%%%%%%%%%%%%%%%%%%%%%%%%%%%%%%
\paragraph*{Existence of Interpolating Kink States:}
%%%%%%%%%%%%%%%%%%%%%%%%%%%%%%%%%%%%%%%%%%%%%%%%%%%%%%%%%%%%%%%%%%%%%%%%%%%%
A criterion 
for the existence of an interpolating kink state $\om\in\6S(\al|\om_1,\om_2)$,
can be obtained by looking at the construction method of \cite{Schl4}.
In our context, we have to select a class of dynamics which are 
equipped with {\em good properties}. Such a selection criterion is 
developed in section 5. We shall show that each 
dynamic of a $P(\phi)_2$-model satisfies this criterion which leads 
to the following result:
%%%%%%%%%%%%%%%%%%%%%%%%%%%%%%%%%%%%%%%%%%%%%%%%%%%%%%%%%%%%%%%%%%%%%%%%%%%
\bthe\label{the1}
If $\al\in\dyn(\6M)$ is a dynamic of a $P(\phi)_2$-model, then for each 
pair of vacuum states $\om_1,\om_2\in\6S_0(\al)$ there exists an 
interpolating kink state $\om\in\6S(\al|\om_1,\om_2)$.
\ethe
%%%%%%%%%%%%%%%%%%%%%%%%%%%%%%%%%%%%%%%%%%%%%%%%%%%%%%%%%%%%%%%%%%%%%%%%%%

We postpone the proof of Theorem \ref{the1} until section 6, since we 
need some further results for preparation.
%%%%%%%%%%%%%%%%%%%%%%%%%%%%%%%%%%%%%%%%%%%%%%%%%%%%%%%%%%%%%%%%%%%%%%%%%%%%

%%%%%%%%%%%%%%%%%%%%%%%%%%%%%%%%%%%%%%%%%%%%%%%%%%%%%%%%%%%%%%%%%%%%%%%%%%
\section{A Criterion for the Existence of an Interpolating Kink-State}
%%%%%%%%%%%%%%%%%%%%%%%%%%%%%%%%%%%%%%%%%%%%%%%%%%%%%%%%%%%%%%%%%%%%%%%%%%

%%%%%%%%%%%%%%%%%%%%%%%%%%%%%%%%%%%%%%%%%%%%%%%%%%%%%%%%%%%%%%%%%%%%%%%%%%%%
\paragraph*{Technical Preliminaries:}
%%%%%%%%%%%%%%%%%%%%%%%%%%%%%%%%%%%%%%%%%%%%%%%%%%%%%%%%%%%%%%%%%%%%%%%%%%%
As mentioned in the introduction, let us 
consider the net which is given by the 
$N$-fold W*-tensor product
$$
\6F_N:\2I\mapsto \6F_N(\2I):=\6M(\2I)^{\olt N}
$$
As usual, we denote by $C^*(\6F_N)$ the C*-algebra which is generated by all 
local algebras $\6F_N(\2I)$.
The permutation group $S_N$ acts by automorphisms on $C^*(\6F_N)$, i.e.:
$$
\sgm\in S_N\mapsto \al_\sgm: a_1\otimes\cdots\otimes a_N\mapsto 
a_{\sgm(1)}\otimes\cdots\otimes a_{\sgm(N)}
$$

\paragraph*{\it Observation:}
We observe from Proposition \ref{pro1} that 
the inclusion of wedge algebras
$$
\6F_N(\2I_{RR})\subset \6F_N(\2I_R)
$$
is split. Moreover, the net $\6F_N$ fulfills Haag duality 
(see \cite{Schl4}), i.e.
$$
\6F_N(\2I^c)=\6F_N(\2I_{LL})\vee\6F_N(\2I_{RR})
$$
where $\2I^c:=\2I\bsl\Sgm$ denotes the complement of $\2I$ in $\Sgm$.
\paragraph*{}

If we interpret $\6F_N$ as a net of field algebra with internal 
symmetry group $S_N$, we can apply the analysis of \cite{Mue1}.

From the observation above, we obtain for each bounded interval 
$\2I$ a unitary representation of the 
permutation group
$$
U_\2I:\sgm\in S_N\mapsto U_\2I(\sgm)\in \6F_N(\2I_R)
$$
which implements the action of the  automorphism group 
$\{\al_\sgm:\sgm\in S_N\}$
on $\6F_N(\2I_{RR})$, i.e.:
$$
\al_\sgm(a)=U_\2I(\sgm) a U_\2I(\sgm)^* 
\ \ \ ; \ \ \ \forall a\in \6F_N(\2I_{RR}) \ \ \ . 
$$
The representations $U_\2I$ can be obtained by using the 
universal localizing map $\Psi_\2I$ (see Section 1). 
The Hilbert-space $\2H_0^{\otimes N}$ carries naturally
a unitary representation $U$ of $S_N$ 
and the representation $U_\2I$ is simply given by 
$$
U_\2I:=\Psi_\2I\circ U:S_N\to\6F_N(\2I_R)\ \ \ .
$$

The adjoint action of $U_\2I(\sgm)$
maps the algebra $\6F_N(\2I_1)$ onto itself 
for \newline 
$\2I_1\supset\2I$ (See \cite{Mue1,Schl4}).
Hence the implementing operator $U_\2I(\sgm)$ 
induces an automorphism 
\beq
\al_\sgm^{\2I}:=\Ad(U_{\2I}(\sgm))
\eeq 
of the algebra $C^*(\6F_N)$. 
Finally, we construct a 
non-local extension of the net $\2I\mapsto \6F_N(\2I)$ (see \cite{Mue1}):
$$
\hat\6F_N:\2I\mapsto\hat\6F_N(\2I):=\6F_N(\2I)\vee U_{\2I}(S_N)
$$

%%%%%%%%%%%%%%%%%%%%%%%%%%%%%%%%%%%%%%%%%%%%%%%%%%%%%%%%%%%%%%%%%%%%%%%%%%%%
\paragraph*{Extendible Dynamics:}
%%%%%%%%%%%%%%%%%%%%%%%%%%%%%%%%%%%%%%%%%%%%%%%%%%%%%%%%%%%%%%%%%%%%%%%%%%%%

We are now prepared to introduce the notion of extendible dynamic. 

%%%%%%%%%%%%%%%%%%%%%%%%%%%%%%%%%%%%%%%%%%%%%%%%%%%%%%%%%%%%%%%%%%%%%%%%%%%%%%
\bdef\label{def5}
Let $\al\in\dyn(\6M)$ be a dynamic of $\6M$. 
We call $\al$ {\em $N$-extendible}
if there is a dynamic $\hat\al^N$ of the extended net $\hat\6F_N$ such that 
$$
\hat\al^N_t|_{C^*(\6F_N)}=\al^N_t:=\al^{\otimes N}_t 
\ \ \ ; \ \ \ \forall t\in\7R
$$
\eef
%%%%%%%%%%%%%%%%%%%%%%%%%%%%%%%%%%%%%%%%%%%%%%%%%%%%%%%%%%%%%%%%%%%%%%%%%%%%%

Here a dynamic of $\hat\6F_N$ is defined in the sense of 
Definition \ref{def1} by replacing the net 
$\6M$ by the non-local extended net $\hat\6F_N$.

%%%%%%%%%%%%%%%%%%%%%%%%%%%%%%%%%%%%%%%%%%%%%%%%%%%%%%%%%%%%%%%%%%%%%%%%%%%%%
\blem\label{lem2}
If a dynamic is 2-extendible, then it is $N$-extendible for each $N\geq 2$.
\elem
%%%%%%%%%%%%%%%%%%%%%%%%%%%%%%%%%%%%%%%%%%%%%%%%%%%%%%%%%%%%%%%%%%%%%%%%%%%%%
\bpr
If we apply the discussion of \cite{Mue1} to our situation, we conclude 
that 
$$
\sum_{\sgm\in S_N} a_\sgm \times u_\sgm \mapsto
\sum_{\sgm\in S_N} a_\sgm U_{\2I}(\sgm) 
$$
is a faithful representation of the crossed-product $\6F_N(\2I)\rtimes S_N$.
Now each permutation can be decomposed into a product of transpositions 
and the result follows.
\epr
%%%%%%%%%%%%%%%%%%%%%%%%%%%%%%%%%%%%%%%%%%%%%%%%%%%%%%%%%%%%%%%%%%%%%%%%%%%%% 

In the sequel, we call a dynamic which is 2-extendible simply {\em extendible}.

Given a bounded interval $\2I$.
We consider for each pair $(\1x,\sgm)\in\7R\times S_N$ the operator 
$$
\gam^\2I_\sgm(\1x)=U_\2I(\sgm)^*\al_\1x(U_\2I(\sgm))=
U_\2I(\sgm)^*U_{\2I+\1x}(\sgm) \ \ \ .
$$ 
The family $\{\gam^\2I_\sgm(\1x);\1x\in\7R\}$
of unitary operators has useful properties which are given in the lemma below.

%%%%%%%%%%%%%%%%%%%%%%%%%%%%%%%%%%%%%%%%%%%%%%%%%%%%%%%%%%%%%%%%%%%%%%%%%%%
\blem\label{lem3}
The family $\{\gam^\2I_\sgm(\1x);\1x\in\7R\}$ 
of unitary operators has the 
properties:
\begin{description}
\item[{\it (1)}]
The map $\gam^\2I_\sgm:\1x\mapsto\gam^\2I_\sgm(\1x)$ is strongly continuous.
\item[{\it (2)}]
For each pair $\1x,\1y\in\7R$ we have:
$\gam^\2I_\sgm(\1x+\1y)=\gam^\2I_\sgm(\1x)\al_\1x\gam^\2I_\sgm(\1y)$
\item[{\it (3)}]
For $\2I=(\1y_1,\1y_2)$ and $\1x>0$, 
$\gam^\2I_\sgm(\1x)$ is contained in $\6F_N(\1y_1,\1x+\1y_2)$, and for 
$\1x<0$, $\gam^\2I_\sgm(\1x)$ is contained in $\6F_N(\1x+\1y_1,\1y_2)$.
\end{description}
\elem
%%%%%%%%%%%%%%%%%%%%%%%%%%%%%%%%%%%%%%%%%%%%%%%%%%%%%%%%%%%%%%%%%%%%%%%%%%%
\bpr
{\it (3)}: For $\1x>0$ the operator $U_{\2I+x}(\sgm)$ is contained in 
$\6F_N(\1x+\1y_1,\infty)$ and implements $\al_\sgm$ on
$\6F_N(\1x+\1y_2,\infty)$. We have now for each  
$a\in\6F_N(-\infty,\1y_1)$ and $a'\in\6F_N(\1x+\1y_2,\infty)$
$$
\begin{array}{l}
\gam^\2I_\sgm(\1x)a\gam^\2I_\sgm(\1x)^*
=U_\2I(\sgm)^*U_{\2I+x}(\sgm)aU_{\2I+x}(\sgm)^*U_\2I(\sgm)
=a
\vs
\gam^\2I_\sgm(\1x)a'\gam^\2I_\sgm(\1x)^*
= U_\2I(\sgm)^*U_{\2I+x}(\sgm)a'U_{\2I+x}(\sgm)^*U_\2I(\sgm)
=\al_\sgm^{-1}\al_\sgm(a')=a'
\end{array}
$$
which implies $\gam^\2I_\sgm(\1x)\in\6F_N(\1y_1,\1x+\1y_2)$. 
The proof for $\1x<0$ works
analogously.

The properties {\it (1)} and {\it (2)} follow directly from the 
construction of $\gam^\2I_\sgm(\1x)$.
\epr
%%%%%%%%%%%%%%%%%%%%%%%%%%%%%%%%%%%%%%%%%%%%%%%%%%%%%%%%%%%%%%%%%%%%%%%%%%%%%%

A one-parameter family, which satisfies the conditions 
{\it (1)} and {\it (2)} in the lemma above, is called a {\em 2-cocycle}
\cite{Rb2,Froh1,Froh3}.

%%%%%%%%%%%%%%%%%%%%%%%%%%%%%%%%%%%%%%%%%%%%%%%%%%%%%%%%%%%%%%%%%%%%%%%%%%%%%%

Let us discuss now the relations between the automorphism 
$\al^\2I_\sgm$ 
and a dynamic $\al\in\dyn(\6M)$.

%%%%%%%%%%%%%%%%%%%%%%%%%%%%%%%%%%%%%%%%%%%%%%%%%%%%%%%%%%%%%%%%%%%%%%%%%%%%%
\blem\label{lem4}
If the dynamic $\al\in\dyn(\6M)$ is extendible, then 
for each $\sgm\in S_N$ the automorphism
$$
(\al^\2I_\sgm)^{-1}\circ\al^N_t\circ \al^\2I_\sgm\circ\al^N_{-t}
$$
of $C^*(\6F_N)$ is inner. 
\elem
%%%%%%%%%%%%%%%%%%%%%%%%%%%%%%%%%%%%%%%%%%%%%%%%%%%%%%%%%%%%%%%%%%%%%%%%%%%%%
\bpr
Since the dynamic $\al$ is extendible there is a dynamic 
$\hat\al^N\in\dyn(\hat\6F_N)$ of the extended net $\hat\6F_N$.
We consider the operator 
$$
\gam^\2I_\sgm(t):=U_\2I(\sgm)^*\hat\al^N_t(U_\2I(\sgm))
$$
and show that it implements the action of the automorphism 
above.
Now we compute for $a\in C^*(\6F_N)$:
\begin{eqnarray*}
\Ad(\gam^\2I_\sgm(t))a= U_\2I(\sgm)^*\hat\al^N_t(U_\2I(\sgm))a
\hat\al^N_t(U_\2I(\sgm)^*)U_\2I(\sgm)
\vs\vs
=
U_\2I(\sgm)^*\hat\al^N_t(U_\2I(\sgm)\al^N_{-t}(a)U_\2I(\sgm)^*)U_\2I(\sgm)
\vs\vs
=
U_\2I(\sgm)^*\hat\al^N_t\biggl(\al^\2I_\sgm(\al^N_{-t}(a))\biggr )U_\2I(\sgm)
\vs\vs
=
U_\2I(\sgm)^*\al^N_t\biggl(\al^\2I_\sgm(\al^N_{-t}(a))\biggr )U_\2I(\sgm)
\vs\vs
=
(\al^\2I_\sgm)^{-1}\biggl(
\al^N_t\biggl(\al^\2I_\sgm(\al^N_{-t}(a))\biggr )\biggr )
\end{eqnarray*}
Using the fact that $\ps(\al)<1$ we can find for each $t\in\7R$ 
a bounded interval $\2I_t=(\1x_1(t),\1x_2(t))$ such that for each
$\1y_1<\1x_1(t)$, for each  $\1y_2>\1x_2(t)$ and for each 
$a\in\6F_N(\1y_1,\1x_1(t))\vee\6F_N(\1x_2(t),\1y_2)$   
we have:
$$
(\al^\2I_\sgm)^{-1}\circ\al^N_t\circ \al^\2I_\sgm\circ\al^N_{-t}(a)=a
$$
This implies that $\gam_\sgm^\2I(t)$ is contained in\newline 
$\6F_N(-\infty,\1x_1(t))'\vee\6F_N(\1x_2(t),\infty)'=\6F_N(\2I_t)$.
\epr
%%%%%%%%%%%%%%%%%%%%%%%%%%%%%%%%%%%%%%%%%%%%%%%%%%%%%%%%%%%%%%%%%%%%%%%%%%%

%%%%%%%%%%%%%%%%%%%%%%%%%%%%%%%%%%%%%%%%%%%%%%%%%%%%%%%%%%%%%%%%%%%%%%%%%%%
\paragraph*{A Criterion for the Existence of Interpolating Kink States:}
%%%%%%%%%%%%%%%%%%%%%%%%%%%%%%%%%%%%%%%%%%%%%%%%%%%%%%%%%%%%%%%%%%%%%%%%%%
Now we are ready to formulate a criterion for the existence of an 
interpolating kink state.

%%%%%%%%%%%%%%%%%%%%%%%%%%%%%%%%%%%%%%%%%%%%%%%%%%%%%%%%%%%%%%%%%%%%%%%%%%%
\bpro\label{pro2}
Let $\al\in\dyn(\6M)$ be an extendible dynamic, then for each pair 
of vacuum states $\om_1,\om_2\in\6S_0(\al)$ 
the state
$$
\om:=\om_1\otimes\om_2\circ\be^\2I|_{C^*(\6M)\otimes\7C\11} 
$$
is an interpolating kink-state, i.e. $\om\in\6S(\al|\om_1,\om_2)$.
\epro
%%%%%%%%%%%%%%%%%%%%%%%%%%%%%%%%%%%%%%%%%%%%%%%%%%%%%%%%%%%%%%%%%%%%%%%%%%%
\bpr
To prove the statement above, we apply the construction scheme 
which is outlined in \cite{Schl4}.
Let us consider the case $N=2$. We have $S_N=\7Z_2=\{1,-1\}$ 
and denote the automorphism with respect to the non-trivial element
by $\be^\2I:=\Ad(U_\2I(-1))$.
By Lemma \ref{lem3} and Lemma \ref{lem4} we conclude that the automorphisms 
$$
\be^\2I\circ\al_{(t,\1x)}\circ\be^\2I\circ\al_{-(t,\1x)}
$$
are inner which implies that 
the representation 
$$
\pi:=\pi_1\otimes\pi_2\circ\be^\2I|_{\6A\otimes\7C\11} 
$$ 
is translationally covariant, i.e.
there exists a unitary strongly-continuous representation of the 
translation group 
$$
U:(t,\1x)\mapsto U(t,\1x)
$$ 
on $\2H_1\otimes\2H_2$ such that: 
$$
\pi\circ\al_{(t,\1x)}=\Ad(U(t,\1x))\circ\pi
$$
Furthermore, it can be shown that the spectrum of the generator of $U$ 
is contained in the closed forward light cone.
If we use the arguments of \cite{Schl4}, we conclude that 
$\pi$ is a cyclic representation which implies that 
$\pi$ is unitarily equivalent to the GNS-representation of $\om$.
Hence $\om$ satisfies the Borchers criterion.

We consider now two bounded intervals $\2I_1\subset (-\infty,\1x)$ and
$\2I_2\subset (\1y,\infty)$.  
We obtain for operators $a_1\in\6M(\2I_1)$ and $a_2\in\6M(\2I_2)$:
$$
\om_\2I(a_1)=\om_1(a_1) \ \ \mbox{  and  } \ \ \om_\2I(a_2)=\om_2(a_2)
$$
Thus we conclude that the state
$\om_\2I$ has the correct interpolation property.
\epr
%%%%%%%%%%%%%%%%%%%%%%%%%%%%%%%%%%%%%%%%%%%%%%%%%%%%%%%%%%%%%%%%%%%%%%%%%%%

%%%%%%%%%%%%%%%%%%%%%%%%%%%%%%%%%%%%%%%%%%%%%%%%%%%%%%%%%%%%%%%%%%%%%%%%%%%
\paragraph*{Multi-Kink States:}
%%%%%%%%%%%%%%%%%%%%%%%%%%%%%%%%%%%%%%%%%%%%%%%%%%%%%%%%%%%%%%%%%%%%%%%%%%%%%

The construction of kink states 
which is described in the proof of Proposition \ref{pro2}
can naturally be generalized to a 
construction of {\em multi-kink states}. We formulate 
this statement in the following Corollary:

%%%%%%%%%%%%%%%%%%%%%%%%%%%%%%%%%%%%%%%%%%%%%%%%%%%%%%%%%%%%%%%%%%%%%%%%%%
\bcor
Let $(\om_1,\cdots,\om_N)\subset\6S_0(\al)$ be a family of vacuum states with 
respect to an extendible dynamic $\al$ and let 
$\2I_1,\cdots,\2I_N$ be bounded intervals, then the 
state 
$$
\om:=\om_1\otimes\cdots\otimes\om_N
\circ\al_{s_N}^{\2I_N}\cdots\al_{s_1}^{\2I_1}|_{\6A\otimes\7C\11}
$$
is an interpolating kink-state which is contained in
$\6S(\al|\om_1,\om_N)$.
\ecor
%%%%%%%%%%%%%%%%%%%%%%%%%%%%%%%%%%%%%%%%%%%%%%%%%%%%%%%%%%%%%%%%%%%%%%%%%%%%
\bpr
We use analogous arguments as in the proof of 
Proposition \ref{pro2}, to conclude that the representation
$$
\pi=\pi_1\otimes\cdots\otimes\pi_N
\circ\al_{s_N}^{\2I_N}\cdots\al_{s_1}^{\2I_1}|_{\6A\otimes\7C\11}
$$
is translationally covariant. 
If we generalize the methods of \cite{Schl4} to the \newline
$N\geq2$ case, 
then we obtain that $\om$ satisfies the Borchers criterion.
   
We consider an 
observable $a\in\6M(\2I)$ with $\2I>\2I_1\cup\cdots\cup\2I_N$ and we obtain
\begin{eqnarray*}
\om(a)=\om_1\otimes\cdots\otimes\om_N
\circ\al_{s_N}^{\2I_N}\cdots\al_{s_1}^{\2I_1}(a\otimes\11)
\vs\vs
=\om_1\otimes\cdots\otimes\om_N
\circ\al_{s_N\cdots s_1}(a\otimes\11)
=\om_N(a) \ \ \ .
\end{eqnarray*}
Analogously we obtain for $a'\in\6M(\2I')$ with 
$\2I'<\2I_1\cup\cdots\cup\2I_N$
$$
\om(a')=\om_1(a')
$$ 
since the automorphisms $\al_{s_j}^{\2I_j}$ act trivially on $\6M(\2I')$.
Thus the correct interpolation property follows immediately.
\epr
%%%%%%%%%%%%%%%%%%%%%%%%%%%%%%%%%%%%%%%%%%%%%%%%%%%%%%%%%%%%%%%%%%%%%%%%%%%%

The state $\om$ can be interpreted as a multi-kink state. 
To motivate this interpretation, we consider a family of intervals 
$(\2I_j=(\1x_j,\1y_j); j=1,\cdots,N)$ such that $\1y_j<\1x_{j+1}$.  
For each operator $a\in\6M(\1y_j,\1x_{j+1})$ we obtain:
\begin{eqnarray*}
\om(a)=\om_1\otimes\cdots\otimes\om_N
\circ\al_{s_N}^{\2I_N}\cdots\al_{s_1}^{\2I_1}(a\otimes\11)
\vs\vs
\om_1\otimes\cdots\otimes\om_N
\circ\al_{s_j}^{\2I_j}\cdots\al_{s_1}^{\2I_1}(a\otimes\11)
\vs\vs
=\om_1\otimes\cdots\otimes\om_N
\circ\al_{s_j\cdots s_1}(a\otimes\11)
\vs\vs
=\om_j(a)
\end{eqnarray*}
Hence the state $\om$ describes a configuration of $N$ kinks, 
where the kink which is localized in $\2I_j$ interpolates the 
vacua $\om_{j-1}$ and $\om_j$. 

%%%%%%%%%%%%%%%%%%%%%%%%%%%%%%%%%%%%%%%%%%%%%%%%%%%%%%%%%%%%%%%%%%%%%%%%%%%
\section{Kink States in $P(\phi)_2$-Models}
%%%%%%%%%%%%%%%%%%%%%%%%%%%%%%%%%%%%%%%%%%%%%%%%%%%%%%%%%%%%%%%%%%%%%%%%%%

Let us consider the dynamic $\al^{P(\phi)}\in\dyn(\6M)$ 
of a $P(\phi)_2$-model. 
As already mentioned, there are familiar $P(\phi)_2$-models for which the set 
$\sec_0(\al^{P(\phi)})$ contains more than one element.  
It is well known that for the $\lam\phi^4_2$-model, the set 
$\sec_0(\al^{\lam \phi^4})$ contains two elements for suitable values of 
the coupling constant $\lam$, i.e.:
$$
\#\sec_0(\al^{\lam\phi^4}) =2 
$$
We shall show that each dynamic of a $P(\phi)_2$-model is extendible.
For this purpose, let us briefly discuss the properties of them.
As described in section 2 the dynamic of a $P(\phi)_2$-model
consists of two parts.
\begin{description}
\item[{\it (1)}]
The first part is given by the free dynamic $\al_0$, with 
propagation speed $\ps(\al_0)=1$, 
$$
\al_{0,t}(a)=e^{ih_0t}ae^{-ih_0t}
$$
which is given by the free Hamiltonian $(h_0,D(h_0))$ 
which is a self-adjoint operator
on the domain $D(h_0)\subset \2H_0$.
\item[{\it (2)}]
The second part is a dynamic $\al_1$ with propagation speed $\ps(\al_1)=0$, 
i.e. it maps each local algebra $\6M(\2I)$ onto itself.
As described in section 2, the interacting part is given by 
a Wick polynomial of the time-zero field $\phi_0$, i.e.
$$
h_1(\2I)=h_1(\chi_\2I)=:P(\phi_0):(\chi_\2I)
$$
where $\chi_\2I$ is a smooth test function which is one on 
$\2I$ and zero on the complement of a 
slightly lager region $\hat\2I\supset\2I$.
The unitary operator $\exp(ith_1(\2I))$ implements the dynamic 
$\al_1$ locally i.e. for each $a\in\6M(\2I)$ we have:
$$
\al_{1,t}(a):=e^{ih_1(\2I)t}ae^{-ih_1(\2I)t} 
$$
\end{description}
%%%%%%%%%%%%%%%%%%%%%%%%%%%%%%%%%%%%%%%%%%%%%%%%%%%%%%%%%%%%%%%%%%%%%%%%%
\bdef\label{def6}
A dynamic $u\in\dyn(\6M)$ of $\6M$ is called {\em ultra local} if there exists 
an operator valued distribution $v:S(\7R)\to L(\2H_0)$ which satisfies
the following properties:
\begin{description}
\item[{\it (1)}]
For each real valued test function $f\in S(\7R)$, 
with $\supp (f)\subset \2I$, the operator $v(f)$ is 
essentially self adjoint on $C^\infty(h_0)=\cap_{n\in\7N}D(h_0^n)$,
affiliated with $\6M(\2I)$ and we have 
$v(f)C^\infty(h_0)\subset C^\infty(h_0)$.
\item[{\it (2)}] 
For each pair of test functions $f_1,f_2\in S(\7R)$, the operators
$v(f_1)$ and $v(f_2)$ commute on $C^\infty(h_0)$.
\item[{\it (3)}]
For each bounded interval $\2I$ and for each operator $a\in\6M(\2I)$,
the dynamic $u$ is implemented by the unitary one parameter group \newline
$\{\exp(itv(\chi)); t\in\7R\}$, i.e.
$$
u_t(a)=\exp(itv(\chi))a\exp(-itv(\chi))
$$
where $\chi\in S(\7R)$ is a positive test function which is one
on $\2I$.
\end{description} 
\eef
%%%%%%%%%%%%%%%%%%%%%%%%%%%%%%%%%%%%%%%%%%%%%%%%%%%%%%%%%%%%%%%%%%%%%%%%%%
It is shown by Glimm and Jaffe \cite{GlJa1}
that the interacting part of
the dynamic of a $P(\phi)_2$-model is ultra local.
Moreover each ultra local dynamic $u$ has propagation speed
$\ps(u)=0$.

The idea is to extend each part of the dynamic separately. Since 
the free part of the dynamic can be extended to the algebra 
of all bounded operators on Fock space $\6B(\2H_0)$, it is clear
that it is extendible for all $N\in\7N$.

%%%%%%%%%%%%%%%%%%%%%%%%%%%%%%%%%%%%%%%%%%%%%%%%%%%%%%%%%%%%%%%%%%%%%%%%%%%
\blem\label{lem5}
Each ultra local dynamic $u\in\dyn(\6M)$ is extendible. 
\elem
%%%%%%%%%%%%%%%%%%%%%%%%%%%%%%%%%%%%%%%%%%%%%%%%%%%%%%%%%%%%%%%%%%%%%%%%%%
\bpr
Let us consider any ultra local dynamic $u\in\dyn(\6M)$ which is given by an
operator-valued distribution $v$ which satisfies the
conditions of the definition above. 
Let $\2I=(a,b)$ be a bounded interval. We write
$$
V(\chi|t):= 
\exp\biggl( itv(\chi)\biggr)
$$
and for the $N$-fold tensor product:
$V_N(\chi|t):=V(\chi|t)^{\otimes N}$.
We consider now for each $\eps >0$ 
test functions $\chi_m,\chi_\eps\in S(\7R)$ such that 
$$
\chi_m(\1x)=\left\{
\begin{array}{ll}
1 & \1x\in (-m,m) \\
0 & \1x\in (-\infty,-m-1)\cup(m+1,\infty)
\end{array} \right.
$$
$$
\chi_\eps(\1x)=\left\{
\begin{array}{ll}
1 & \1x\in (a-\eps,b+\eps ) \\
0 & \1x\in (-\infty,a-2\eps)\cup(b+2\eps,\infty)
\end{array} \right.
$$
For $m>b+\eps$ and $-m<a-\eps$, 
there are test functions $\chi^\pm_{m,\eps}\in S(\7R)$ with 
$$
\begin{array}{l}
\supp(\chi^-_{m,\eps})\subset (-m-1, a-\eps)
\vs 
\supp(\chi^+_{m,\eps})\subset (b+\eps,m+1)
\vs
\chi_m-\chi_\eps=\chi^+_{m,\eps}+\chi^-_{m,\eps} 
\end{array}
$$
In the sequel, we use the following notation:
$$
V(m|t):=V(\chi_m|t) \ \ \ ; \ \ \ V(\eps|t):=V(\chi_\eps|t)
\ \ \ ; \ \ \ V_\pm(m,\eps|t):=V(\chi^\pm _{m,\eps}|t)
$$
Since we have $[v(\chi_1),v(\chi_2)]=0$ for any pair of test functions
$\chi_1,\chi_2\in S(\7R)$, we obtain for each $\eps>0$:
\beq
V_N(m|t)= V_N(\eps|t)V_{N,-}(m,\eps|t)V_{N,+}(m,\eps|t)
\eeq
If we use the fact, that $V_{N,\pm}(m,\eps|t)$ is 
$\al_\sgm$-invariant, for each $\sgm\in S_N$, we obtain
\beq
\Ad(V_N(m|t))U_{(a,b)}(\sgm)= \Ad(V_N(\eps|t))U_{(a,b)}(\sgm)
\eeq
which depends only of the localization interval $(a,b)$.
Hence we conclude that for $a\in\hat\6F_N(a,b)$ and for $-m<a<b<m$ 
$$
\hat u_t^N(a):=\Ad(V_N(m|t))a
$$
defines a dynamic of $\hat\6F_N$ whose restriction to $\6F_N$
is $u^{\otimes N}$. 
\epr
%%%%%%%%%%%%%%%%%%%%%%%%%%%%%%%%%%%%%%%%%%%%%%%%%%%%%%%%%%%%%%%%%%%%%%%%%%%%

If $\hat\al_0^N$ denotes the natural extension of the free dynamic
to $\hat\6F_N$ and let $\hat u$ be the extension of an ultra local 
dynamic then, by using the Trotter product, we conclude that the 
dynamic 
$$
\hat\al:=\hat \al_0^N\times \hat u^N
$$
is an extension of the dynamic $(\al_0\times u)^{\otimes N}$ to $\hat\6F_N$.
This leads to the following result:
%%%%%%%%%%%%%%%%%%%%%%%%%%%%%%%%%%%%%%%%%%%%%%%%%%%%%%%%%%%%%%%%%%%%%%%%%%%
\bpro\label{pro3}
Each dynamic of a $P(\phi)_2$-model is extendible.
\epro
%%%%%%%%%%%%%%%%%%%%%%%%%%%%%%%%%%%%%%%%%%%%%%%%%%%%%%%%%%%%%%%%%%%%%%%%%%%
\bpr
The statement follows from Lemma \ref{lem5} 
and due to the fact that each dynamic 
of a $P(\phi)_2$-model is a Trotter product of the free dynamic $\al_0$
and an ultra local dynamic $\al_1$. 
\epr
%%%%%%%%%%%%%%%%%%%%%%%%%%%%%%%%%%%%%%%%%%%%%%%%%%%%%%%%%%%%%%%%%%%%%%%%%%%

\paragraph*{Proof of Theorem \ref{the1}:}
The statement of Theorem \ref{the1} is 
an immediate consequence which is the formulated in the corollary below.
%%%%%%%%%%%%%%%%%%%%%%%%%%%%%%%%%%%%%%%%%%%%%%%%%%%%%%%%%%%%%%%%%%%%%%%%%%
\bcor
Let $\al\in\dyn(\6M)$ be a dynamic of a $P(\phi)_2$-model, then 
for each pair of vacuum states $\om_1,\om_2\in\6S_0(\al)$ there
exists an interpolating kink state $\om\in\6S(\al|\om_1,\om_2)$.
\ecor
%%%%%%%%%%%%%%%%%%%%%%%%%%%%%%%%%%%%%%%%%%%%%%%%%%%%%%%%%%%%%%%%%%%%%%%%%%%%
\bpr
By Proposition \ref{pro3} 
each dynamic of a $P(\phi)_2$-model is extendible and we can apply 
Proposition \ref{pro2} which implies the result.
\epr
%%%%%%%%%%%%%%%%%%%%%%%%%%%%%%%%%%%%%%%%%%%%%%%%%%%%%%%%%%%%%%%%%%%%%%%%%%%

%%%%%%%%%%%%%%%%%%%%%%%%%%%%%%%%%%%%%%%%%%%%%%%%%%%%%%%%%%%%%%%%%%%%%%%%%%%%
\section{Conclusion and Outlook}
%%%%%%%%%%%%%%%%%%%%%%%%%%%%%%%%%%%%%%%%%%%%%%%%%%%%%%%%%%%%%%%%%%%%%%%%%%%%
We have seen that for each pair of vacuum states which belong to 
a dynamic of a $P(\phi)_2$-model, there exists an interpolating kink state. 
This result can be obtained by a generalization 
of the methods which are used by J. Fr\"ohlich in \cite{Froh1,Froh3}.
The assumption that the interpolated 
vacua are related by an internal 
symmetry transformation is not needed for the application of our 
construction scheme. 
Furthermore, the construction is 
independent of specific properties of a model and uses only 
the extendibility condition of its dynamic.

Familiar examples of super symmetric models (Wess-Zumino models), 
which are described in \cite{JanWeit},   
have more then one vacuum state and their dynamics
consist of a $P(\phi)_2$-like and a $\mbox{Yukawa}_2$-like part.
We conjecture that there are also kink states in 
$\mbox{Yukawa}_2$-like models.
By using the construction of the dynamic of the 
$\mbox{Yukawa}_2$ model, which is discussed by Glimm and Jaffe \cite{GlJa1},
we can use similar technics as above, to show that the dynamic of  
a $\mbox{Yukawa}_2$-like model is extendible.   
Therefore, we belief that our results can also be applied to 
this class of models.

%%%%%%%%%%%%%%%%%%%%%%%%%%%%%%%%%%%%%%%%%%%%%%%%%%%%%%%%%%%%%%%%%%%%%%%%%%%% 
\subsubsection*{{\it Acknowledgment:}}
I am grateful to Prof. K. Fredenhagen for 
supporting this investigation with many ideas.
Thanks are also 
due to my colleagues in Hamburg for careful reading.
%%%%%%%%%%%%%%%%%%%%%%%%%%%%%%%%%%%%%%%%%%%%%%%%%%%%%%%%%%%%%%%%%%%%%%%%%%%%%

%%%%%%%%%%%%%%%%%%%%%%%%%%%%%%%%%%%%%%%%%%%%%%%%%%%%%%%%%%%%%%%%%%%%%%%%%%
\begin{appendix}
\section{Appendix}
%%%%%%%%%%%%%%%%%%%%%%%%%%%%%%%%%%%%%%%%%%%%%%%%%%%%%%%%%%%%%%%%%%%%%%%%%
\subsection*{Remarks on the Split Property for Massive Free Scalar Fields}
%%%%%%%%%%%%%%%%%%%%%%%%%%%%%%%%%%%%%%%%%%%%%%%%%%%%%%%%%%%%%%%%%%%%%%%%%%%
We are going to prove the generalization of Proposition \ref{pro1} to 
any number of spatial dimensions. 

%%%%%%%%%%%%%%%%%%%%%%%%%%%%%%%%%%%%%%%%%%%%%%%%%%%%%%%%%%%%%%%%%%%%%%%%%%%
\paragraph*{Preliminaries:}
%%%%%%%%%%%%%%%%%%%%%%%%%%%%%%%%%%%%%%%%%%%%%%%%%%%%%%%%%%%%%%%%%%%%%%%%%%%
For our purpose it is convenient to work with the self-dual  
CCR-algebra (in the sense of Araki). Therefore, we need some 
technical definitions. 
%%%%%%%%%%%%%%%%%%%%%%%%%%%%%%%%%%%%%%%%%%%%%%%%%%%%%%%%%%%%%%%%%%%%%%%%%%%%%
\bdef\label{def7}
For the vector space $K=S(\7R^d)\oplus S(\7R^d)$, we denote by $\Gamma$ 
the complex conjugation in $K$, $\Gamma f=\bar f$. Moreover, we 
introduce the following sesquilinear form $\gamma$ on $K$:
\beq
\gamma(f,g)=\biggl ( f , \pmatrix{0&-i\cr
                                  i&0\cr} g\biggr )
\eeq
where $(\cdot,\cdot)$ denotes the ordinary scalar-product in 
$L_2(\7R^d)\oplus L_2(\7R^d)$.
The {\em self-dual CCR-algebra}
$\6A(K,\gam,\Gam)$ is the *-algebra 
which is generated by the set of symbols
$\{ b(f): f\in K\}$ modulo the following relations:
\begin{description}
\item[{\it (1)}]
The map $b:f\in K\mapsto b(f)\in \6A(K,\gam,\Gam)$ is linear.  
\item[{\it (2)}]
We have the following *-relation: $b(f)^*=b(\Gam f)$.
\item[{\it (3)}]
We have the commutator relation
$[b(f)^*,b(g)]=\gam (f,g)\11$.
\end{description}
For a region $G\subset \7R^d$ we consider the CCR-algebra
$\6A(G):=\6A(K(G),\gam,\Gam)$ where $K(G)$ is defined by
$K(G):=\mu^{1/2}S(G)\oplus\mu^{-1/2}S(G)$. Here $\mu$ is the 
pseudo differential operator which is given by kernel
\beq
\mu(\1x-\1y):=\int \8d\1p \ (\1p^2+m^2)^{1/2} \ e^{i\1p(\1x-\1y)} 
\eeq
as described in section 1. 
\eef
%%%%%%%%%%%%%%%%%%%%%%%%%%%%%%%%%%%%%%%%%%%%%%%%%%%%%%%%%%%%%%%%%%%%%%%%%%%%%

We define now the vacuum functional $\om_0$ on $\6A(K,\gam,\Gam)$ by
$$
\om_0(b(f)^*b(g)):= 1/2 \gam(f,g) \ \ \ .
$$
where the functions $f,g$ are contained in $K$.
The GNS-representation of $\om_0$ is unitarily equivalent to the 
representation $\pi_0$ which is given by
$$
b(f)\mapsto \pi_0(b(f)):=
{1\over 2}\biggl (a^*(f_1)+a(f_1)\biggr)
+{1\over 2i}\biggl (a^*(f_2)-a(f_2) \biggr) \ \ \ .
$$
Each test function $f\in K(G)$ can be written of the form 
$$
f=\mu^{-1/2}f_1\oplus\mu^{1/2}f_2
$$
with test functions $f_1,f_2\in S(G)$ and we obtain
$$
\pi_0(b(f))=B(f_1\oplus f_2)
$$
where $B$ denotes the operator valued distribution which 
is given in section 2. 

%%%%%%%%%%%%%%%%%%%%%%%%%%%%%%%%%%%%%%%%%%%%%%%%%%%%%%%%%%%%%%%%%%%%%%%%%%
\paragraph*{Product States:}
%%%%%%%%%%%%%%%%%%%%%%%%%%%%%%%%%%%%%%%%%%%%%%%%%%%%%%%%%%%%%%%%%%%%%%%%%%% 
Let us consider now two regions $G_1,G_2$ with non vanishing 
distance. In the sequel we write $G:=G_1\cup G_2$ for their union. 

We denote by $\6A(G_1)\vee\6A(G_2)$ the algebra 
which is given by all finite sums
$\sum a_nb_n$ with $a_n\in \6A(G_1)$ and $b_n\in\6A(G_2)$.
Since $G_1$ and $G_2$ have non vanishing distance we conclude that
$\6A(G_1)\vee\6A(G_2)=\6A(G)$.
We define now a {\em product} state $\om$ on $\6A(G)$
by 
\beq
\om(\sum a_nb_n):=\sum \om_0(a_n)\om_0(b_n) \ \ \ .
\eeq
Clearly since $\om_0$ is quasi-free, $\om$ is also a quasi-free state 
on $\6A(G)$.

We are now interested in a criterion which give us the 
possibility to decide for which regions $G_1,G_2$ with non vanishing distance 
the GNS-representations with respect to the 
states $\om$ and $\om_0$ are unitarily equivalent
on $\6A(G)$.

We are going to use a criterion which is proven 
by H. Araki \cite{Ark1}.
To formulate this criterion, let us consider the 
following two scalar products on the space $K(G_1\cup G_2)$:
\begin{description}
\item[{\it (1)}]
$
(f,g)_0:=\om_0(b(f)^*b(g))+\om_0(b(\Gam f)^*b(\Gam g))
$
\item[{\it (2)}]
$
(f,g)_p:=\om(b(f)^*b(g))+\om(b(\Gam f)^*b(\Gam g))
$
\end{description}

Here $\om$ is the product state, induced by $\om_0$.
The completion of $K(G)$ with respect to the norm 
$||\cdot||_0=(\cdot,\cdot)_0$ (resp. $||\cdot||_p=(\cdot,\cdot)_p$)
is denoted by $K(G)_0$ (resp. $K(G)_p$).

Moreover, denote by $s_0$ (resp. $s_p$) a positive operator, bounded 
by $1$, with the property $(f,s_0g)_0=\om_0(b(f)^*b(g))$
(resp. $(f,s_pg)_p=\om(b(f)^*b(g))$).

\subparagraph{Criterion:}
The GNS-representations with respect to $\om_0$ and $\om$ 
are unitarily equivalent if the following conditions hold:
\begin{description}
\item[{\it (1)}]
The values $0,1/2$ are not eigenvalues of $s_0$ (resp. $s_p$) in 
$K(G)_0$ (resp. $K(G)_p$).
\item[{\it (2)}]
The norms $||\cdot||_0$ and $||\cdot||_p$ are equivalent on $K(G)$.
\item[{\it (3)}]
The following operators are of Hilbert-Schmidt class in $K(G)_0=K(G)_p$:
$$
(s_0-s_p)(\11-2s_0)^{-1} \ \ \mbox{ and } \ \ 
(s_0(\11-s_0))^{1/2}-(s_p(\11-s_p))^{1/2}
$$
\end{description}

The following analysis can be done in complete analogy to those of D. Buchholz 
\cite{Bu1} who has proven that $\om$ and $\om_0$ are unitarily
equivalent on $\6A(G)$, in the case where $G_1=O_1$ is a compact region and 
$G_2=O_2$ the complement of a slightly larger compact region in $\7R^3$.
The only argument in this analysis which depends on the spatial dimension
is contained in the proof of condition {\it (2)} (\cite[Lemma 3.2]{Bu1}). 
The necessary generalization
is given in the next paragraph.   

If one carries through the analysis of \cite{Bu1}, we obtain the following
criterion:
Consider two regions $\hat G_j\supset G_j \ \ ; \ \ j=1,2$ such that 
$\hat G_1$ and $\hat G_2$ have also non vanishing distance and let 
$\chi_{G_1},\chi_{G_2}$ be two $C^\infty$-functions with 
$\supp(\chi_{G_j})\subset \hat G_j$ and 
$\chi_{G_j}(\1x)=1$ for $\1x\in G_j$. Then we obtain:

\bpro\label{pro4}
The states $\om$ and $\om_0$ are unitarily equivalent on $\6A(G)$ 
if the integral-kernel
\beq
\chi_{G_1}(\1x)\mu(\1x-\1y)\chi_{G_2}(\1y)
\eeq
is an element of $S(\7R^{2d})$.
\epro
%%%%%%%%%%%%%%%%%%%%%%%%%%%%%%%%%%%%%%%%%%%%%%%%%%%%%%%%%%%%%%%%%%%%%%%%%%%%%
\paragraph*{Equivalence of Norms:}
%%%%%%%%%%%%%%%%%%%%%%%%%%%%%%%%%%%%%%%%%%%%%%%%%%%%%%%%%%%%%%%%%%%%%%%%%%%%%
For convenience, we cite now the proof of \cite[Lemma 3.2]{Bu1} 
by making the necessary 
changes to show that the result is independent of the spatial dimension. 

\blem\label{lem6}
Let $(G_1,G_2)$ be any pair of regions with non-vanishing distance, then 
the norms $||\cdot||_0$ and $||\cdot||_p$ are equivalent on $K(G_1\cup G_2)$.
\elem 

\bpr
Let $t>0$ be the distance between $G_1$ and $G_2$. Moreover, let $s$,
be a function in $S$ with support in $B_d(t/2)$ and Fourier transform 
$\hat s$, such that $\hat s(\1p)\geq 0$ for all $p\in\7R^d$. 
Clearly, a function with 
these properties exists and can be obtained by using the convolution 
theorem. 
Hence there are constants $c>a>0$ such that 
$c>(\1p^2+m^2)^{1/2}(\hat s(\1p)+a)\geq a >0$. This implies 
\beq
\begin{array}{l}
|(\1p^2+m^2)^{-1/2}-c^{-1}(\hat s(\1p)+a)|\leq ac^{-1}(\1p^2+m^2)^{-1/2}
\vs
|(\1p^2+m^2)^{1/2}-c^{-1}(\1p^2+m^2)(\hat s(\1p)+a)|
\leq ac^{-1}(\1p^2+m^2)^{-1/2}
\end{array}
\eeq
We consider now the following operators which are diagonal in momentum space:
\beq
\begin{array}{l}
w_1(\1p)=c^{-1}(\hat s(\1p)+a)
\vs
w_2(\1p)=c^{-1}(\1p^2+m^2)(\hat s(\1p)+a)
\end{array}
\eeq
For any element $g\in S(G_2)$ one has
\beq
\begin{array}{l}
(w_1g)(\1x)=c^{-1}(s*g(\1x)+ag(\1x))
\vs
(w_2g)(\1x)=c^{-1}(\pa_\al\pa^\al+m^2)(s*g+ag)(\1x) \ \ \ , \ \ \ \al=1,2,3
\end{array}
\eeq
and hence $\supp w_jg \cap G_1=\emptyset$. 
Thus one gets $(f,w_jg)=0$ for each $f\in K(G_1)$ and each $g\in K(G_2)$.
Now we compute:
\beq
\begin{array}{l}
|(f,\mu^{-1}g)|=|(f,\mu^{-1}g-w_1g)|
\vs
\leq \int \8d\1p \  
|(\1p^2+m^2)^{-1/2}-c^{-1}(\hat s(\1p)+a)||\hat f(\1p)||\hat g(\1p)|
\vs
\leq \ ac^{-1} \int \8d\1p \ (\1p^2+m^2)^{-1/2}|\hat f(\1p)||\hat g(\1p)|
\vs
\leq \ ac^{-1} (f,\mu^{-1}f)^{1/2}(g,\mu^{-1}g)^{1/2}
\end{array}
\eeq
Analogously we obtain the estimate
$|(f,\mu g)|\leq \ ac^{-1} (f,\mu f)^{1/2}(g,\mu g)^{1/2}$.
Keeping in mind that $ac^{-1}<1$, 
the equivalence of the norms $||\cdot||_0$ and $||\cdot||_p$ can be obtained 
by using the same arguments as in \cite{Bu1}.
\epr
%%%%%%%%%%%%%%%%%%%%%%%%%%%%%%%%%%%%%%%%%%%%%%%%%%%%%%%%%%%%%%%%%%%%%%%%%%%%%%
\paragraph*{Application of the Criterion:}
%%%%%%%%%%%%%%%%%%%%%%%%%%%%%%%%%%%%%%%%%%%%%%%%%%%%%%%%%%%%%%%%%%%%%%%%%%%%%
In this paragraph, we discuss the application of Proposition \ref{pro4} 
with respect to the possible cases for $G_1$ and $G_2$. 

Denote by $S(\7R^d;0)$ the space of functions $f$ such that 
$\chi_G f\in S(\7R^d)$ for each open set $G$ which does not contain 
the point $\1x=0$. Here $\chi_G\in S(\7R^d)$ denotes the 
smoothed characteristic function of a region $G$.

It turns out that 
the problem can be reduced to the following question:

Let $f$ be a function in $S(\7R^d;0)$. For which pairs   
of regions $G_1,G_2\subset\7R^d$ is the function
\beq
f_{(G_1,G_2)}:(\1x,\1y)\mapsto \chi_{G_1}(\1x)f(\1x-\1y)\chi_{G_2}(\1y)
\eeq
contained in $S(\7R^{2d})$ ?
 
Clearly since $f$ may be singular at $\1x=0$,  
one has to require that $G_1$ and $G_2$ have non vanishing 
distance. 

\bdef 
A pair of regions $G_1,G_2\subset\7R^d$ 
with non vanishing distance is called {\em admissible} 
if there exists a constant $k>0$ such that for each $r>0$ the set 
$$
G(r):=
\{(\1x_1,\1x_2)|\1x_1\in G_1 ,\1x_2\in G_2 \  ; \  \1x_1-\1x_2\in B_d(r)\}
$$
is contained in $B_{2d}(kr)$, where $B_d(r)$ denotes the closed ball
in $\7R^d$ with radius $r$.
\eef
 
\blem\label{lem7} 
If $(G_1,G_2)$ is a pair of regions in $\7R^d$ 
witch is admissible, then the function 
$f_{(G_1,G_2)}$ is contained in $S(\7R^{2d})$.
\elem

\bpr
Since the pair $(G_1,G_2)$ is admissible, the region 
$G(k^{-1}r):=
\{(\1x,\1y)|\1x\in G_1 ,\1y\in G_2 \  ; \  \1x-\1y\in B_d(k^{-1}r)\}$ 
is contained
in the closed ball $B_{2d}(r)$ for a constant $k>0$.  This implies 
that for each $m\in\7N$ one has 
\beq
\begin{array}{r}
|\chi_{G_1}(\1x)f(\1x-\1y)\chi_{G_2}(\1y)|< {\rm const.}\cdot |\1x-\1y|^{-m}
\vs 
\leq {\rm const.}\cdot k^m r^{-m} \leq {\rm const.}\cdot |(\1x,\1y)|^{-m} \ \ \ .
\end{array}
\eeq
Hence we conclude that $f_{(G_1,G_2)}$ is of fast decrease and thus contained 
in $S(\7R^{2d})$.
\epr
 
\bcor
If the pair of regions $(G_1,G_2)$ is admissible, then 
the states $\om_0$ and $\om$ are unitarily equivalent on $\6A(G)$.
\ecor

\bpr 
The function 
\beq
f:\1x\in\7R^d\backslash\{0\}\mapsto f(\1x)=\int \8d\1p \ 
(\1p^2+m^2)^{1/2} \ e^{i\1p\1x}  
\eeq
is contained in $S(\7R^d;0)$. An application of Proposition \ref{pro4} 
and 
Lemma \ref{lem7} implies the result.
\epr

Let us now discuss the cases for witch the pair $(G_1,G_2)$ is admissible.
To carry through this analysis, we have to give a few more definitions.
Let $e\in\7R^d$ be a vector of unit length and $s\in (0,1)$, then 
we define the convex cone $C(e,s):=\7R_+\cdot (B_d(s)+e)$. The complement 
of $C(e,s)$ in $\7R^d$ is denoted by $C'(e,s)$.

\blem\label{lem8} 
Let $s_1,s_2\in (0,1)$ with $s_1<s_2$ and $e$ a unit vector, then 
for each $\epsilon >0$ the pair 
$(C(e,s_1)+\epsilon e,C'(-e,s_2))$ is admissible.
\elem

\bpr
Let us consider the set $C(e,s_2)\backslash C(e,s_1)=C(e,s_2,s_1)$.
For $s_2>s_1$, there exists a convex cone $C(e',s_3)$ which is contained in
$C(e,s_2,s_1)$. Hence for each $\1x\in \pa C(e,s_1)$ exists $r>0$, such that
$B_d(r)+\1x\subset C(e,s_2)$. Moreover, we have the following relation between
$\1x$ and $r$:
\beq
|\1x|\geq \sin(\varphi_2-\varphi_1)^{-1} \cdot r
\eeq
Here $\varphi_j=\arcsin(s_j)$ is the opening angle of $C(e,s_j)$.
We set $t:=\sin(\varphi_2-\varphi_1)^{-1}$ and conclude for
each $\1x\in B_d(tr)'\cap C(e,s_1)$ 
\beq
B_d(r)\subset C(-e,s_2)+\1x \ \ \ .
\eeq
Hence for each $\1x\in B_d(tr)'\cap C(e,s_1)$ there 
is no $\1y\in C'(-e,s_2)$ such that $\1x+\1y\in B_d(r)$. Since for each 
$\epsilon>0$ the set   
$C(e,s_1)+\epsilon e$ is contained in $C(e,s_1)$, we obtain that 
\beq
G(r):=\{(\1x,\1y)|\1x\in C(e,s_1)+\epsilon e,\1y\in C'(-e,s_2) \  
; \  \1x+\1y\in B_d(r)\}
\eeq
is contained in $B_d(tr)\times C'(-e,s_2)$. 
On the other hand, for each $r>0$ there exists $\1y\in \pa C(-e,s_2)$ such that 
$B_d(r)\cap C(e,s_1)=\emptyset$. We have the following 
relation for $\1y$ and $r$:
\beq
|\1y|\geq \sin(\varphi_2-\varphi_1)^{-1} \cdot r
\eeq
Thus with the same argument as above we conclude finally that there 
exists a constant $k>0$, such that
\beq
G(r)\subset B_d(tr)\times B_d(tr)\subset B_{2d}(kr) 
\eeq
which implies the result. 
\epr

We see that for $d>1$ the arguments in the proof of Lemma \ref{lem8} fails
for cones with the same opening angle, i.e.
the pair $(C(e,s)+\epsilon e,C'(-e,s))$ is {\em not} admissible.

On the other hand, for $d=1$ the pair $((-\infty,0],[\epsilon,\infty))$
is indeed admissible. 
%%%%%%%%%%%%%%%%%%%%%%%%%%%%%%%%%%%%%%%%%%%%%%%%%%%%%%%%%%%%%%%%%%%%%%%%%%%%%
\paragraph*{The Split Property:}
%%%%%%%%%%%%%%%%%%%%%%%%%%%%%%%%%%%%%%%%%%%%%%%%%%%%%%%%%%%%%%%%%%%%%%%%%%%%
To discuss the split property, we briefly describe the construction 
of the local v.Neumann algebras for the free massive scalar field
in the vacuum representation.
Denote by $(\2H_0,\pi_0,\Om_0)$ the GNS-triple 
of $\om_0$. 
We define for each $f\in K_\Gam:=\{g\in K:\Gam g=g\}$, the field operator 
$b_0(f):=\pi_0(b(f))$ which is essentially 
self-adjoint on $\pi_0(\6A(K,\gam,\Gam))\Om_0$. For a region $G\subset\7R^d$
we denote by $\6M(G)$ the v.Neumann algebra which is given by
$\6M(G):=\{ e^{i\pi_0(b(f))}: f\in K_\Gam(G) \}''$, where $''$ denotes the 
double commutant in $\2B(\2H_0)$. 

Let us consider a pair of admissible regions $(G_1,G_2)$, then 
by Corollary 3.1 we know that the vacuum state $\om_0$ and 
its induced product state $\om$ are unitarily equivalent on
$\6A(G_1\cup G_2)$. Hence the product state $\om$ induces a normal
state on $\6M(G_1)\vee\6M(G_2)$ which is given by a vector
$\eta\in \2H_0$, where $\eta$ is cyclic for $\6M(G_1)\vee\6M(G_2)$.
Thus we have for $a_1\in \6M(G_1)$ and $a_2\in \6M(G_2)$
\beq
\<\eta,a_1a_2\eta\>=\<\Om_0,a_1\Om_0\> \ \<\Om_0,a_2\Om_0\> 
\eeq

By standard arguments \cite{Bu1}, we conclude that for a pair of admissible 
regions $(G_1,G_2)$ the inclusion 
\beq
\6M(G_1)'\subset\6M(G_2)
\eeq
is a split inclusion.

\subparagraph{Example:} 
We close the appendix by discussing the $1+1$-dimensional
case briefly. We consider the regions $(0,\infty)$ and $(-\infty,0)$.
For $\1x\in (0,\infty)$ the pair $((\1x,\infty),(-\infty,0))$ 
is admissible (see Lemma \ref{lem8}).
Keeping in mind that the net of the free field $\2I\mapsto\6M(\2I)$
satisfies wedge duality we obtain that the inclusion 
\beq 
\6M(\1x,\infty)\subset\6M(0,\infty)
\eeq
is standard split. Hence the massive free scalar field in $1+1$ dimensions 
satisfies the split property for wedge regions.
%%%%%%%%%%%%%%%%%%%%%%%%%%%%%%%%%%%%%%%%%%%%%%%%%%%%%%%%%%%%%%%%%%%%%%%%%%%%%%
\end{appendix}
%%%%%%%%%%%%%%%%%%%%%%%%%%%%%%%%%%%%%%%%%%%%%%%%%%%%%%%%%%%%%%%%%%%%%%%%%%%%%
%%%%%%%%%%%%%%%%%%%%%%%%%%%%%%%%%%%%%%%%%%%%%%%%%%%%%%%%%%%%%%%%%%%%%%%%%%%%%
%%%%%%%%%%%%%%%%%%%%%%%%%%%%%%%%%%%%%%%%%%%%%%%%%%%%%%%%%%%%%%%%%%%%%%%%%%%%

%%%%%%%%%%%%%%%%%%%%%%%%%%%%%%%%%%%%%%%%%%%%%%%%%%%%%%%%%%%%%%%%%%%%%%%%%%%%%%%
\end{document}